\documentclass[
reprint,
superscriptaddress,
amsmath,
amssymb,
aps,
physrev,
prb,
longbibliography,
]{revtex4-2}

\usepackage{graphicx}
\usepackage{dcolumn}
\usepackage{bm}
\usepackage{xcolor}

\usepackage{hyperref}

\begin{document}

\title{\textbf{
Superfluidity in the spin-1/2 XY model with power-law interactions}}

\author{Muhammad Shaeer Moeed}%
\affiliation{Department of Physics and Astronomy, University of Waterloo, Waterloo, Ontario, N2L 3G1, Canada}
\affiliation{Perimeter Institute for Theoretical Physics, Waterloo, Ontario, N2L 2Y5, Canada}
\author{Costanza Pennaforti}%
\affiliation{Department of Physics and Astronomy, University of Waterloo, Waterloo, Ontario, N2L 3G1, Canada}
\affiliation{Perimeter Institute for Theoretical Physics, Waterloo, Ontario, N2L 2Y5, Canada}
\author{Adrian Del Maestro}%
\affiliation{Department of Physics and Astronomy, University of Tennessee, Knoxville, Tennessee, 37996, USA}
\affiliation{Institute for Advanced Materials and Manufacturing,
University of Tennessee, Knoxville, Tennessee, 37996, USA}
\affiliation{Min H. Kao Department of Electrical Engineering and Computer Science, University of Tennessee, Knoxville, Tennessee, 37996, USA}
\author{Roger G. Melko}%
\affiliation{Department of Physics and Astronomy, University of Waterloo, Waterloo, Ontario, N2L 3G1, Canada}
\affiliation{Perimeter Institute for Theoretical Physics, Waterloo, Ontario, N2L 2Y5, Canada}
\email{rmelko@perimeterinstitute.ca}

\date{\today}

\begin{abstract}
In trapped-ion quantum simulators, effective spin-1/2 XY interactions can be engineered via laser-induced coupling between internal atomic states and collective phonon modes. In the simplest one-dimensional ($1d$) traps, these interactions decay as a power-law with distance $1/r^{\alpha}$, with a tunable exponent $\alpha$.
For small $\alpha$, the resulting long-range $1d$ XY model exhibits continuous symmetry breaking, in marked contrast to its nearest neighbor counterpart.
In this paper, we examine this  model near the phase transition at $\alpha_c$ from the lens of the spin stiffness, or superfluid density.
We develop a stochastic series expansion (SSE) quantum Monte Carlo (QMC) simulation and
a generalized winding number estimator to measure the superfluid density in the presence of power-law interactions, which we test against exact diagonalization for small lattice sizes.
Our results show how conventional superfluidity in the $1d$ XY model is enhanced in the long-range interacting regime. This is observed as a diverging superfluid density as $\alpha \rightarrow 0$ in the thermodynamic limit, which we show is consistent with linear spin-wave theory.
Finally, we define a normalized superfluid density estimator that clearly distinguishes the short, medium, and long-range interacting regimes, providing a novel QMC probe of the critical value $\alpha_c$.

\end{abstract}

\maketitle

\section{\label{sec:Intro} Introduction}

In recent years, advances in the development of atomic control have paved the road for quantum simulations of engineered spin models \cite{islam2011onset, kim2010quantum, monroe2021programmable}. 
In particular, trapped ion devices can be used to natively implement one-dimensional ($1d$) models with $U(1)$ symmetry, such as XY and XXZ spin-$1/2$ chains, with power-law decaying interactions \cite{monroe2021programmable}. Experimentally, the distribution of couplings $J_{ij}$ between lattice sites $i$ and $j$ closely follows a $1/r^{\alpha}$ interaction, where $r$ is the distance between the corresponding sites and $\alpha$ is the interaction strength exponent which can be continuously varied using trap and laser control parameters \cite{richerme2014non, islam2013emergence}. In some modern architectures, $\alpha$ can be tuned between $0$ and $3$ with modest control requirements \cite{teoh2020machine, lewis2023ion}. 
Characterizing the phase diagrams of these systems, consequently, is of immediate practical relevance in the context of near-term quantum computers. 

The nearest-neighbor variant of the $1d$ spin-$1/2$ XY model is a cornerstone of low-dimensional condensed matter physics. The ground state of this system admits quasi-long range order, as exhibited by the power-law decay of its in-plane correlations \cite{mermin_wagner, lsm_theorem}, and is well-understood theoretically since it can be mapped to a theory of free-fermions \cite{jordan1928paulische}. Moreover, efficient density matrix renormalization group (DMRG) \cite{schollwock2011density, schollwock2005density, verstraete2023density} and quantum Monte Carlo (QMC) \cite{Sandvik2010Review, melko2013SSEReview, ceperley1995path} techniques have been developed to numerically probe ground state and finite temperature quantities of interest in nearest-neighbor spin models, including the XY model.

In the presence of power-law interactions, renormalization group (RG) power series expansions indicate that for $\alpha > 2+d$, where $d$ is the spatial dimension, the critical exponents of the resulting spin models are identical to those of the nearest neighbor models \cite{fisher_critical_exponents}. 
Indeed,
bosonization theory suggests that the short range part of the $1d$ XY Hamiltonian dominates the behavior of the ground state for $\alpha > 3$ \cite{maghrebi2017continuous}. 
In contrast, for interactions that decay sufficiently slowly, the long-range terms become relevant and the corresponding $1d$ XY model admits a novel continuous symmetry breaking (CSB) phase, absent in the nearest-neighbor model \cite{maghrebi2017continuous, edmond_SCHA, linked_cluster_long_range_xy, TommasoLSWLRXY}. This phase is characterized by a continuously varying dynamical critical exponent $z = (\alpha - 1)/2$ which decreases as $\alpha$ decreases and saturates at $\alpha = 1$ to $z=0$. Recent linear spin wave (LSW) theory analysis suggests that there are two distinct regimes associated with this CSB phase, the long-range limit characterized by $\alpha < 1$, and the medium-range regime corresponding to intermediate values of the interaction strength exponent: $\alpha \in (1,3)$ \cite{TommasoLSWLRXY}. 

For $\alpha = 0$, the system converges to a variant of the Lipkin-Meshkov-Glick (LMG) model which can be solved exactly \cite{lipkin1965validity, meshkov1965validity, glick1965validity}. Since the Hamiltonian is super-extensive for $\alpha \leq 1$, Kac normalization \cite{kac1963van, schuckert2025observation} is typically used to restore the extensivity of the energy in this regime. In the thermodynamic limit, the ground state properties of this LMG model converge to mean field theory (MFT) results. Moreover, LSW theory indicates that the ground state of the entire long-range regime is given by the mean field ground state $\left(|E_0\rangle = \otimes_i^N |+\rangle_i\right)$ in the thermodynamic limit \cite{TommasoLSWLRXY}. It should also be noted that while LSW predicts a phase boundary of $\alpha_c = 3.0$, perturbative RG and DMRG calculations suggest that the critical point $\alpha_c$ is slightly smaller than $3.0$ \cite{maghrebi2017continuous}.

Motivated by these results, in this work, we develop a stochastic series expansion (SSE) \cite{sandvik1992generalization, sandvik2002directedloops, KPS_Review, sandvik_SSE_Ising} QMC algorithm for the power-law spin-$1/2$ XY chain and use it to investigate the behavior of the superfluid density (or spin stiffness) as a function of the interaction strength exponent $\alpha$. We supplement our analysis with mean field and LSW calculations, and also discuss the behavior of the superfluid density in the two exactly solvable limits ($\alpha \rightarrow 0$ and $\alpha \rightarrow \infty$). We find that the long range hopping of bosons enhances the superfluid density in the limit of $\alpha \rightarrow 0$ which manifests as a diverging superfluid density. This is also supported by Luttinger Liquid (LL) theory \cite{haldane_luttinger_liquids, quantum_phys_1d_Giamarchi, DelMaestro:2010ib, eggel2011dynamical, Dupuis_2020} results which suggest that in the presence of power-law interactions, the superfluid density diverges for $\alpha < 3$ \cite{dupuis_2024}, as reflected in our QMC calculations. 

We also define a normalized superfluid density estimator to control the diverging contribution in the spin stiffness estimator for the superfluid density. This normalized superfluid density suggests three distinct regimes in $\alpha$; the long-range limit ($\alpha < 1$) which is well-described by MFT, the short-range model ($\alpha > 3$) in which the superfluid density monotonically increases until it achieves its maximum in the nearest-neighbor limit and finally, a medium-range regime bounded by $\alpha \approx 1$ and $\alpha \approx 3$. Here, the normalized superfluid density monotonically decreases until it achieves its minimum at $\alpha \approx 2.7$. The existence of these regimes is consistent with the findings of Ref. \onlinecite{TommasoLSWLRXY} which used DMRG and LSW to analyze the ground state of the same system. Since the diverging component is explicitly normalized away in this estimator, it is better able to distinguish the different $\alpha$ dependent regimes in the power-law XY chain than the spin stiffness, which is dominated by the divergence for $\alpha < \alpha_c$.  

This paper is organized as follows: in Sec. \ref{sec:LR_XY_Model}, we discuss the power-law decay XY model Hamiltonian as well as our convention for imposing periodic and twisted boundary conditions required to compute the superfluid density. In Sec. \ref{sec:exact_limits}, we discuss the exact analytical solutions corresponding to $\alpha = 0$ and the limit $\alpha \rightarrow \infty$. Sec. \ref{sec:qmc} is dedicated to the development of our QMC approach which is validated using exact diagonalization (ED) studies, the exact limits as well as the MFT and LSW results discussed in Sec. \ref{sec:approx_sols}. Finally, we discuss the superfluid behavior of the power-law decay XY model in Sec. \ref{sec:stiffness_enhancement} and conclude with some future research directions in Sec. \ref{sec:conclusion}. 

\section{XY Model with power-law interactions \label{sec:LR_XY_Model}}

\subsection{\label{sec:XY_Model} Hamiltonian}

In this work, we study the $1d$ spin-$1/2$ XY model with power-law decay interactions in the presence of periodic boundaries \cite{maghrebi2017continuous}: 
\begin{equation}
    H = -J \sum_{i < j}^{N} \frac{1}{r_{ij}^\alpha} \left(S_i^{x} S_j^{x} + S_{i}^{y} S_{j}^{y}\right) . \label{sum_pairs_H}
\end{equation}

For our computational studies in this paper, we will always set the energy scale to $J=1$. 
We define sites on the $1d$ lattice to have spacing $a=1$, so that the distance measure in units of this lattice spacing is $r/a=r$.
To implement periodic boundaries on a chain, $r_{ij}$ between sites  $i$ and $j$ is defined as the minimum winding distance (see Fig. \ref{fig:long_range_boson_hopping}):
\begin{equation}
    r_{ij} = \text{min} \{|i-j|, N - |i-j|\} .
\end{equation}

We can equivalently express the Hamiltonian using spin-$1/2$ raising and lowering operators $S_i^{\pm} = S_i^x \pm i S_i^y$ using the relation $\left(S_i^{x} S_j^{x} + S_{i}^{y} S_{j}^{y}\right) = (S_i^+ S_j^- + S_i^-S_j^+)/2$. Moreover, leveraging the translational invariance of the chain, we can re-express the sum over all pairs as follows:
\begin{equation}
    H = -\frac{J}{2} \sum_{i=0}^{N-1} \sum_{r=1}^{(N-1)/2}\frac{1}{r^\alpha} (S_i^+ S_{i+r}^- + S_i^-S_{i+r}^+) . \label{sum_distances_H}
\end{equation}
Here, we have used the fact that $S_{N+k}^\pm = S_{k}^\pm$ and assumed an odd number of spins $N$ in the $1d$ lattice. We will use the form in Eq. \eqref{sum_pairs_H} to construct our QMC algorithm and the form in Eq. \eqref{sum_distances_H} will prove to be more useful for our theoretical analysis. 

\subsection{\label{sec:stiffness} Superfluid Density}

\begin{figure}
    \centering
    \includegraphics[width=1.0\linewidth]{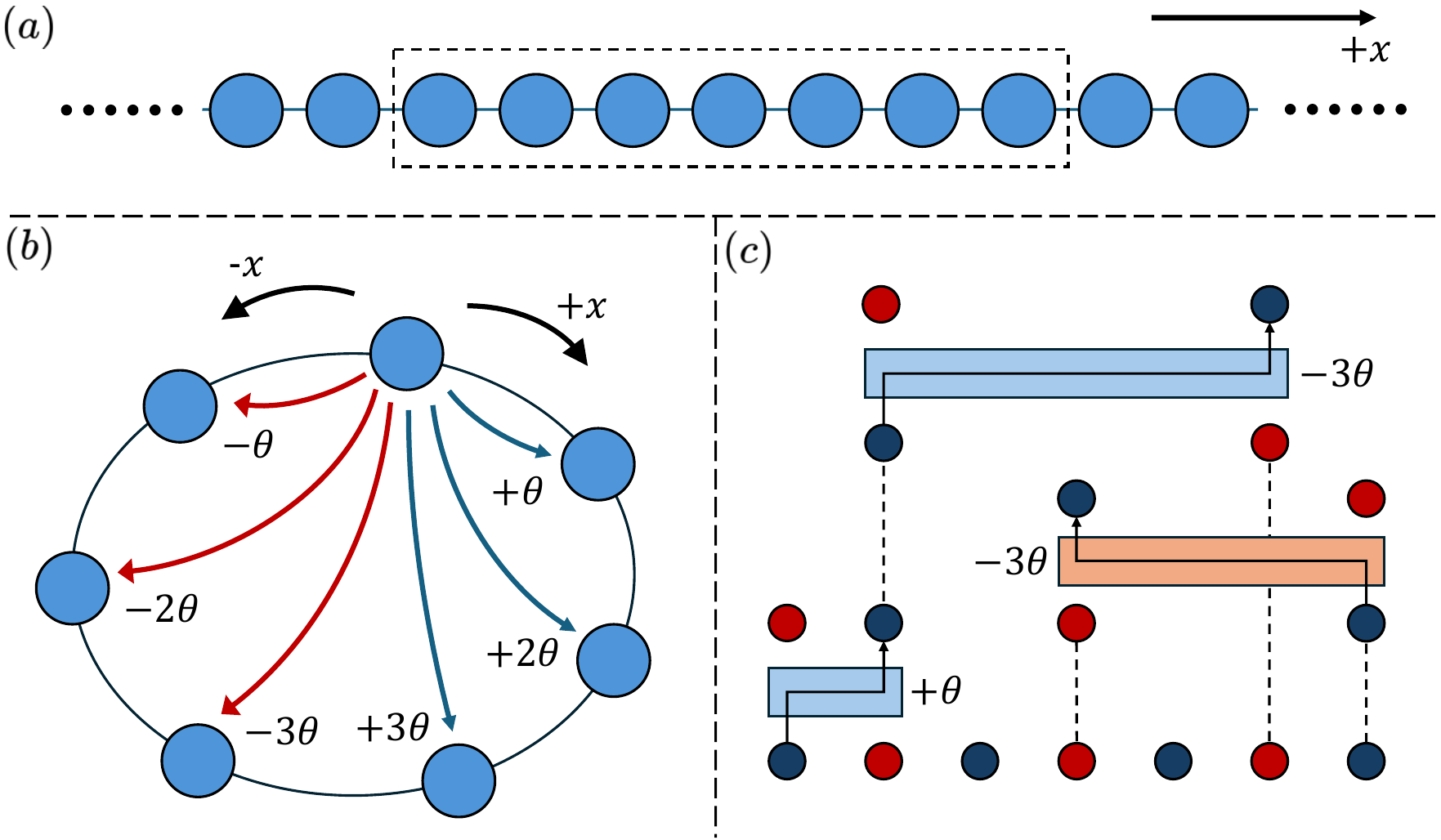}
    \caption{$1d$ lattice of sites with periodic boundaries. Panel $(a)$ exhibits an infinite chain of spin-1/2 particles or equivalently, the sites between which the bosons hop in the hardcore boson model. The simulation cell is highlighted with the dotted line box surrounding the seven sites in the middle. Panel $(b)$ shows our convention for imposing periodic and twisted boundaries on the simulation cell in $(a)$. Panel $(c)$ displays the SSE operator loop diagram associated with this convention.}
    \label{fig:long_range_boson_hopping}
\end{figure}

Note that Eq. \eqref{sum_distances_H} readily admits a mapping to hardcore bosons under the transformation: $S_i^+ \rightarrow b_i^\dagger$, $S_i^ - \rightarrow b_i$, $S_i^z \rightarrow b_i^\dagger b_i$ - 1/2 which gives \cite{melko2005supersolid}:
\begin{equation}
    H = -\frac{J}{2} \sum_{i=0}^{N-1} \sum_{r=1}^{(N-1)/2}\frac{1}{r^\alpha} (b_i^\dagger b_{i+r} + b_{i}b_{i+r}^\dagger) . \label{hardcore_boson_hopping}
\end{equation}

Superfluid phenomena in the hardcore boson hopping model has been studied extensively in the nearest-neighbor XY model, which is recovered by our Hamiltonian in the limit of $\alpha \rightarrow \infty$ ($\lim_{\alpha \rightarrow \infty} 1/r^{\alpha} = \delta_{r,1}$). In particular, it can been shown analytically that this system admits a non-zero superfluid density at $T=0$ in $1d$ (see Sec. \ref{sec:fermion_sol}) \cite{laflorencie2004scaling, laflorencie2001finite}. Moreover, superfluid phenomena in nearest-neighbor $2d$ square and triangular lattices have been the subject of a multitude of QMC studies in past work \cite{melko2004aspect, melko2005supersolid}. 

The superfluid density can be related to the response of a system under a boundary phase twist distributed linearly over all sites in the lattice \cite{sandvik1997finite, laflorencie2004scaling}. This transforms the bosonic raising and lowering operators as: $b_j \rightarrow e^{-i 2 \pi j\phi/N} b_j$ and $b_j^{\dagger} \rightarrow e^{i 2\pi j \phi/N} b_j^{\dagger}$. For power-law decay interactions, the phase twisted Hamiltonian then becomes:
\begin{equation}
    H = -\frac{J}{2} \sum_{j=1}^{N-1}\sum_{r=1}^{(N-1)/2} \frac{1}{r^{\alpha}} (e^{-i r \theta} b_j^{\dagger} b_{j+r} + e^{ir \theta} b_j b_{j+r}^{\dagger})\, . \label{twisted_long_range}
\end{equation}
Here, we have defined $\theta = 2\pi \phi/N$ and assumed an odd number of spins on the chain. For an even number of spins on a ring, the twist can be imposed by choosing either a positive or a negative phase for the bond connecting sites $i$ and $i + N/2$ for any $i \in \{1,...,N\}$. Then, since the Hamiltonian has to be Hermitian, a hardcore boson hopping from $i$ to $i - N/2$ yields a phase in the opposite direction. Because the distinction between having an even or odd number of spins in the system should not affect the phase diagram in the limit of $N \rightarrow \infty$, we will restrict ourselves to odd $N$ in this work. The spin stiffness estimator of superfluid density at temperature $T$ is then given by the second derivative of the free energy per site \cite{fisher1973helicity, sandvik1997finite}:
\begin{equation}
    \rho_s(T) = \frac{1}{N}\frac{\partial^2 F}{\partial \theta^2} \biggr|_{\theta=0}. \label{helicity_modulus}
\end{equation}

Fig. \ref{fig:long_range_boson_hopping} exhibits our convention for imposing the twisted boundary conditions defined by Eq. \eqref{twisted_long_range}. As shown in panel (a) and discussed above, we consider a simulation cell with an odd number of sites, the size of which is increased to suppress finite size effects. For this schematic, the simulation cell size is $N=7$. Periodic and twisted boundaries are imposed by placing all of the spins in the simulation cell on a ring. A boson hopping clockwise along a bond with distance less than $N/2$ accrues a positive phase proportional to the linear distance between the participating sites and, vice versa for a boson hopping anti-clockwise. 

For QMC, it is useful to translate this into an operator loop diagram \cite{sandvik2002directedloops}, an example of which is shown in panel $(c)$. A boson hopping between sites $1$ and $2$ yields a phase of $+\theta$. The second operator (colored orange) is oriented opposite to the first and has a bond distance $r_b = 3$ which implies a negative phase according to Eq. \eqref{twisted_long_range}. The last operator corresponds to $b_2 b_6^\dagger$ which implies a bond distance of $4 > (N-1)/2 = 3$. This corresponds to the term $b_6^{\dagger}b_{9}$ in Eq. \eqref{twisted_long_range} and is therefore, associated with a negative phase.  

The low energy behavior of the nearest-neighbor XY model in $1d$ is also often described using LL theory \cite{haldane_luttinger_liquids, quantum_phys_1d_Giamarchi, DelMaestro:2010ib, eggel2011dynamical}. The effective LL Hamiltonian is defined in terms of bosonic fields $\theta(x)$ and $\phi(x)$ as follows: 
\begin{equation}
    H_{LL} = \frac{v}{2\pi} \int dx \left[ K_{LL} (\partial_x \theta )^2 + \frac{1}{K_{LL}} (\partial_x \phi )^2\right] .
\end{equation}
The model $H_{LL}$ describes the low energy properties of a quantum fluid in $1d$ with compressibility $\kappa = K_{LL}/\pi v$ and sound velocity $v$. $K_{LL}$ is typically referred to as the Luttinger parameter and depends on the microscopic details. The superfluid density in the LL description is simply given by $\tilde{\rho}_s = K_{LL} v/\pi$. For the nearest-neighbor XY model, it can be shown that $K_{LL} = 1$ and $v = J$ which yields $\tilde{\rho}_s = J/\pi$ \cite{quantum_phys_1d_Giamarchi}. 

This LL formalism can also be extended to power-law interactions in the lattice model \cite{Dupuis_2020, dupuis_2024}. In this case, the effective low energy LL action takes the same form as the nearest-neighbor model. 
However, both the Luttinger parameter $K_{LL}$ and the velocity $v$ depend on the momentum $k$ in this construction and diverge in the low-momentum regime as $\sim |k|^{(\alpha - 3)/2}$ for $\alpha < 3$ \cite{dupuis_2024}. This suggests a diverging superfluid density $\tilde{\rho}_s(k \rightarrow 0) = K_{LL}(k) v(k)/\pi \sim |k|^{(\alpha - 3)}$ for the ground state of the long and medium range regimes of the XY model with power-law interactions. 

In the preceding discussion, we have introduced two notions of superfluidity, the helicity modulus in Eq. \eqref{helicity_modulus}, and the coefficient of an effective long-wavelength action: $\tilde{\rho}_s = K_{LL} v$. In $3d$, these coincide at all temperatures in the thermodynamic limit \cite{two_defs_superfluidity, melko2004aspect}. However, for $1d$ and $2d$ systems, the order of limits: $T \rightarrow 0$ and $N \rightarrow \infty$, determines the convergence of the helicity modulus to the long wavelength superfluid response \cite{melko2004aspect, criteria_superfluid}. In particular, in $1d$, if the limit $N \rightarrow \infty$ is taken before the $T \rightarrow 0$ limit, the helicity modulus vanishes. In contrast, if the order of these limits is reversed, the helicity modulus converges to the long-wavelength definition of the superfluid density \cite{two_defs_superfluidity, superfluidity_1d_bose_hubbard, giamarchi_persistent}. 

For the rest of this work, we will use the helicity modulus definition of the superfluid density, since we are interested in the ground state ($T \rightarrow 0$) limit, and because this quantity is readily accessible via the winding number estimator in QMC. Operationally, this competition between the two limits will present as a crossover in our QMC calculations. For each $N$, there will be a critical temperature $T_c$ above which the helicity modulus will converge to $0$, and for $T < T_c$, we will recover the ground state $\rho_s$. 

\begin{figure*}
    \centering
    \includegraphics[width=0.49\linewidth]{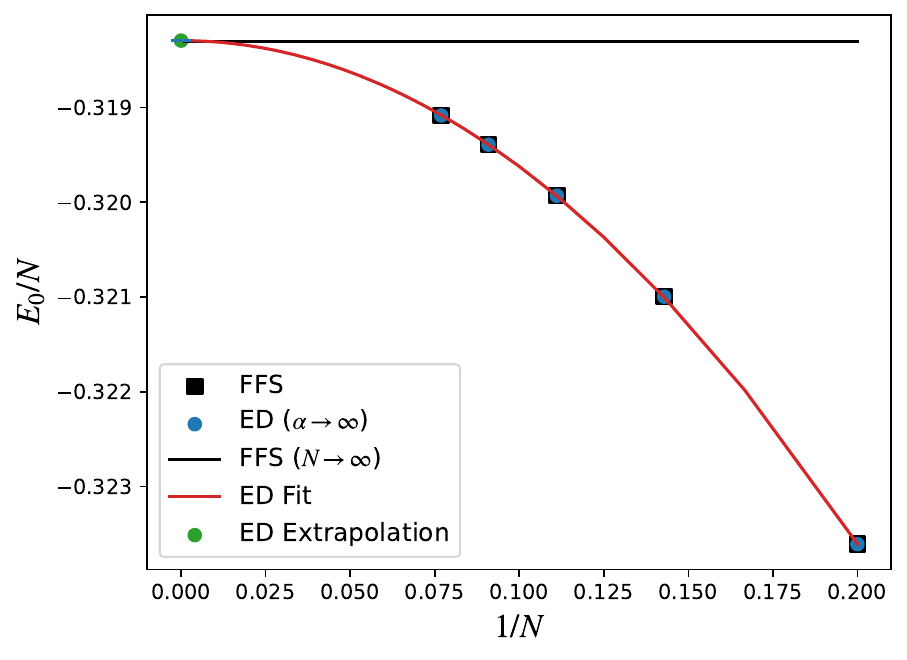}
    \includegraphics[width=0.48\linewidth]{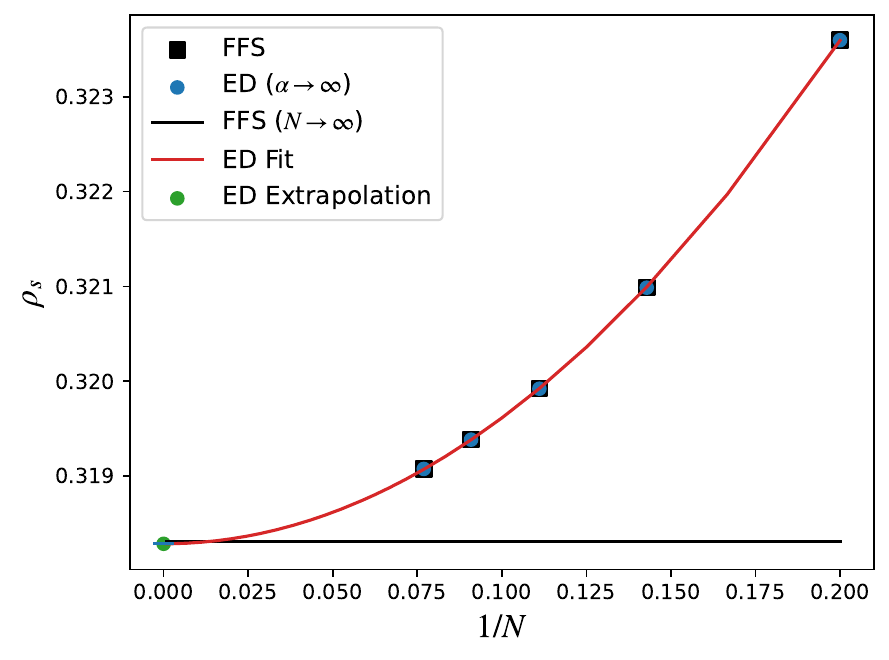}
    \caption{ED results for $\alpha = 50$ (nearest-neighbor limit) compared to the FFS. The black line shows the FSS ($N \rightarrow \infty$) limit and the red line shows the ED results fit. The left panel pertains to $E_0/N$ and the right panel corresponds to $\rho_s$.}
    \label{fig:stiffness_ff_vs_ed}
\end{figure*}

\section{Exact Limits \label{sec:exact_limits}}

\subsection{\label{sec:fermion_sol} Nearest-Neighbor Limit}

The nearest neighbor XY model can be solved exactly in $1d$ using the Jordan-Wigner transform \cite{bialonczyk2021exact, jordan1928paulische}. We will use this to validate our ED and QMC calculations in the limit of $\alpha \rightarrow \infty$. To compute the superfluid density in this case, we need to construct the free-fermion solution in the presence of a phase twist which yields the following Hamiltonian: 
\begin{equation}
    H(\theta) = -\frac{J}{2} \sum_{j=0}^{N-1} (e^{i \theta}S_{j}^+S_{j+1}^- + e^{-i \theta}S_{j}^- S_{j+1}^+).
\end{equation}

The free-fermion solution can then be readily obtained via the Jordan-Wigner transformation (as in the case of periodic boundaries):
\begin{equation}
    S_j^+ = e^{i \pi \sum_{k=1}^{j-1} a_k^\dagger a_k} a_j^\dagger, \ \ S_j^- = e^{-i \pi \sum_{k=1}^{j-1} a_k^\dagger a_k} a_j .
\end{equation}
The string operators $e^{ \pm i \pi \sum_{k=1}^{j-1}a_k^\dagger a_k}$ ensure that the correct spin-$1/2$ commutation relations are recovered between different sites from the fermionic anti-commutation relations.  The resulting Hamiltonian is as follows:
\begin{equation}
    H(\theta) = -\frac{J}{2} \sum_{j=0}^{N-1} ( e^{i\theta} a_j^{\dagger} a_{j+1} + e^{-i \theta}a_{j+1}^{\dagger} a_j).
\end{equation}
This can be diagonalized via a Fourier transform ($a_j = 1/\sqrt{N}\sum_{k=0}^{N-1} e^{-i2 \pi jk/N} f_k$) to yield \cite{laflorencie2001finite, laflorencie2004scaling}:
\begin{equation}
    H(\theta) = -J \sum_{k=0}^{N-1} \epsilon_k(\theta) f_j^\dagger f_j, \ \ \epsilon_k = \cos(2\pi k/N - \theta).
\end{equation}

For $T=0$, the free energy is given by the ground state energy which can be readily obtained by populating all modes associated with a negative eigenvalue. This gives $E_0 = \sum_{\Omega} \epsilon_k$ with $\Omega = \{k : \epsilon_k \geq 0\}$. The superfluid density $\rho_s$ can then be approximated using finite differences computationally. In the limit of $N \rightarrow \infty$, it can be shown analytically that the free-fermion solution (FFS) for $\rho_s$ tends to $J/\pi$ (as expected from LL theory), and the ground state energy $E_0$ tends to $-J/\pi$ with a $\mathcal{O}(1/N^2)$ finite size correction \cite{laflorencie2004scaling}. 

We use the FFS discussed in this section to validate our computational ED approach. Fig \ref{fig:stiffness_ff_vs_ed} shows the ED results for $\alpha = 50$ and $N \in \{5,7,9,11,13\}$. The finite differences have been calculated using $J=1$ and $\theta = 0.02$. For all values of $N$, since $\alpha$ is large, we see good agreement with FFS. The ED results are fit linearly in $1/N^2$: $f(1/N^2) = a/N^2 + b$. As expected, the fit extrapolation to $N \rightarrow \infty$ (given by $b$) converges to the predicted FFS value of $-1/\pi$ for $E_0/N$ and $1/\pi$ for $\rho_s$.

\subsection{ \label{sec:lmg} Lipkin-Meshkov-Glick Limit}

It is also instructive to interrogate the opposite limit: $\alpha = 0$. This yields a chain with constant interaction strengths for each bond. The resulting Hamiltonian is a variant of the LMG model first introduced in the context of nuclear physics \cite{lipkin1965validity, meshkov1965validity, glick1965validity}. The infinite range Hamiltonian of interest is: 
\begin{gather}
    H = -\frac{J}{2}\sum_{i \neq j} (S_i^x S_j^x + S_i^y S_j^y). \label{alpha_0_untwisted_H}
\end{gather}
Here, we have symmetrized the sum using the fact that $J_{ij} = J_{ji}$ and $S_{j}^{\gamma}S_{i}^{\gamma} = S_{i}^{\gamma} S_{j}^{\gamma}$ for $\gamma \in \{x,y\}$. This can be diagonalized exactly using the collective spin operators \cite{LMG2008} (see Appendix \ref{sec:appendix_A_lmg_energy}): 
\begin{equation}
    S_\gamma = \sum_{i=1}^N S_i^\gamma, \ \ \ \gamma \in \{x,y,z\}.
\end{equation}

The ground state energy of this Hamiltonian is given by: 
\begin{equation}
    E_0 = -\frac{J}{8} (N^2 - 1) \label{energy_gs_lmg}
\end{equation}
To determine $\rho_s$, we need to analyze the twisted Hamiltonian: 
\begin{gather}
    H = -\frac{J}{2} \sum_{i \neq j} \biggl ( \cos(r_{ij} \theta) (S_{i}^x S_j^x + S_i^y S_j^y) \nonumber \\ + \sin(r_{ij} \theta) (S_{i}^x S_{j}^x - S_{i}^y S_{j}^y) \biggr). \label{twisted_lgm_H}
\end{gather}
The second term with coefficient $\sin(r_{ij} \theta)$ is the current density $j_b$, which has a trivial expectation value in the ground state \cite{sandvik1997finite}. Therefore, it suffices to only consider the first term in Eq. \eqref{twisted_lgm_H}. We can approximate $\rho_s$ using finite differences as follows: 
\begin{equation}
    \rho_s = \lim_{\theta \rightarrow 0} \frac{2}{N} \frac{\langle H(\theta) \rangle  - \langle H(0) \rangle }{\theta^2}. \label{rho_s_finite_differences_def}
\end{equation}

Note that the collective spin operator can not be directly used here as for the untwisted case because of the bond-dependent coupling $\cos(r_{ij} \theta)$. However, since we are interested in the limit of $\theta \rightarrow 0$, we can restrict to $\theta^2 \ll (N-1)^2 \Rightarrow \cos(r_{ij} \theta) \approx 1 - r_{ij}^2 \theta^2/2$ and analyze the effect of the twist perturbatively. This gives for the superfluid density (see Appendix \ref{sec:appendix_A_lmg_stiffness}): 
\begin{equation}
    \rho_s = \frac{J(N+1)}{8} \left (\frac{N^2 - 1}{12} \right) \label{lmg_stiffness}
\end{equation}

\section{\label{sec:qmc} Quantum Monte Carlo}

We will use QMC to interrogate larger system sizes. To this end, we adapt the deterministic directed loop SSE algorithm for the nearest-neighbor XY model \cite{sandvik2002directedloops, Sandvik2010Review} to power-law decay interactions. SSE proceeds by expressing the partition function $Z = \text{Tr}(e^{-\beta H})$ using a Taylor expansion of the imaginary time propagator $e^{-\beta H}$: 
\begin{equation}
    Z = \sum_{\sigma} \sum_{n=0}^{\infty} \frac{(-\beta)^{n}}{n!} \langle \sigma | H^n | \sigma \rangle.
\end{equation}

Here, $\beta = 1/T$ is the inverse temperature and we have set $k_B = 1$. The ket $|\sigma\rangle$ here represents an $N$-body state: $|\sigma_1,...,\sigma_N\rangle$, where $|\sigma_k\rangle \in \{|0\rangle,|1\rangle\}$ is an eigenstate of $S_z$ for each $k \in \{1,...,N\}$. To construct configurations amenable to importance sampling, we re-express $H$ as a sum over the links (denoted $b = (i,j)$ with $i<j$) connecting different sites in the chain, and add a constant offset, as follows \cite{sandvik2019Review2}:
\begin{gather}
    H' = -J \sum_{b=1}^{N_b} (H_{1,b} + H_{2,b}), \\
    H_{1,b} = \frac{1}{2 r_{b}^\alpha}, \\ H_{2,b} = \frac{1}{2 r_{b}^{\alpha}} \left( S_{i(b)}^{+} S_{j(b)}^{-} + S_{i(b)}^{-} S_{j(b)}^{+} \right).\label{Hamiltonian_pieces}
\end{gather}

The total number of links is $N_b = N (N-1)/2$. Clearly, $H' = -JC + H$ where $C = \sum_b 1/2r_b^\alpha$. Since a constant offset does not affect the partition function, we can use $H'$ to construct the algorithm. Now, taking the $n$th power of $H$ yields a product of sums which can be expressed as a sum over all permutations $S_n = \{(a_1,b_1),...,(a_n,b_n)\}$ with $a_k \in \{1,2\}$ and $b_k \in \{1,...,N_b\}$. This gives for the partition function: 
\begin{equation}
    Z = \sum_{\sigma} \sum_{n=0}^{\infty} \sum_{S_n} \frac{(J\beta)^{n}}{n!} \langle \sigma | \prod_{k=1}^{n} H_{a_k,b_k} \left |\sigma \right \rangle.
\end{equation}

Imposing a cutoff in the expansion direction $n_{max} = M$ (which is set dynamically during equilibration), we can introduce identity fill in operators $H_{0,b} = I$ and expand the set of possible values of $a$ to include $0$ \cite{sandvik2019Review2}. Then, combinatorial re-arrangement of the above yields \cite{melko2013SSEReview}: 
\begin{equation}
    Z = \sum_{\{\sigma\}} \sum_{S_M} \frac{(J\beta)^{n} (M-n)!}{M!} \prod_{k=1}^{M} \langle \sigma_k | H_{a_k,b_k} \left |\sigma_{k+1} \right \rangle.
\end{equation}

Note that we have inserted $M-1$ resolutions of identity between the $M$ operators in each weight, and collected the resulting summations into the sum over $\{\sigma\} = \sigma_1, ..., \sigma_M$. To retain the periodicity of the trace, we impose $\sigma_1 = \sigma_M$. It is also worth noting here that because of Eq. \eqref{Hamiltonian_pieces}, the many body states $|\sigma_k\rangle$ and $|\sigma_{k+1}\rangle$ can either be the same or differ by a single flipped pair of spins. Since the paths need to close, the number of flips attached to any site in a contributing configuration must be even \cite{melko2013SSEReview, Sandvik2010Review}. 

\begin{figure}
    \centering
    \includegraphics[width=1.0\linewidth]{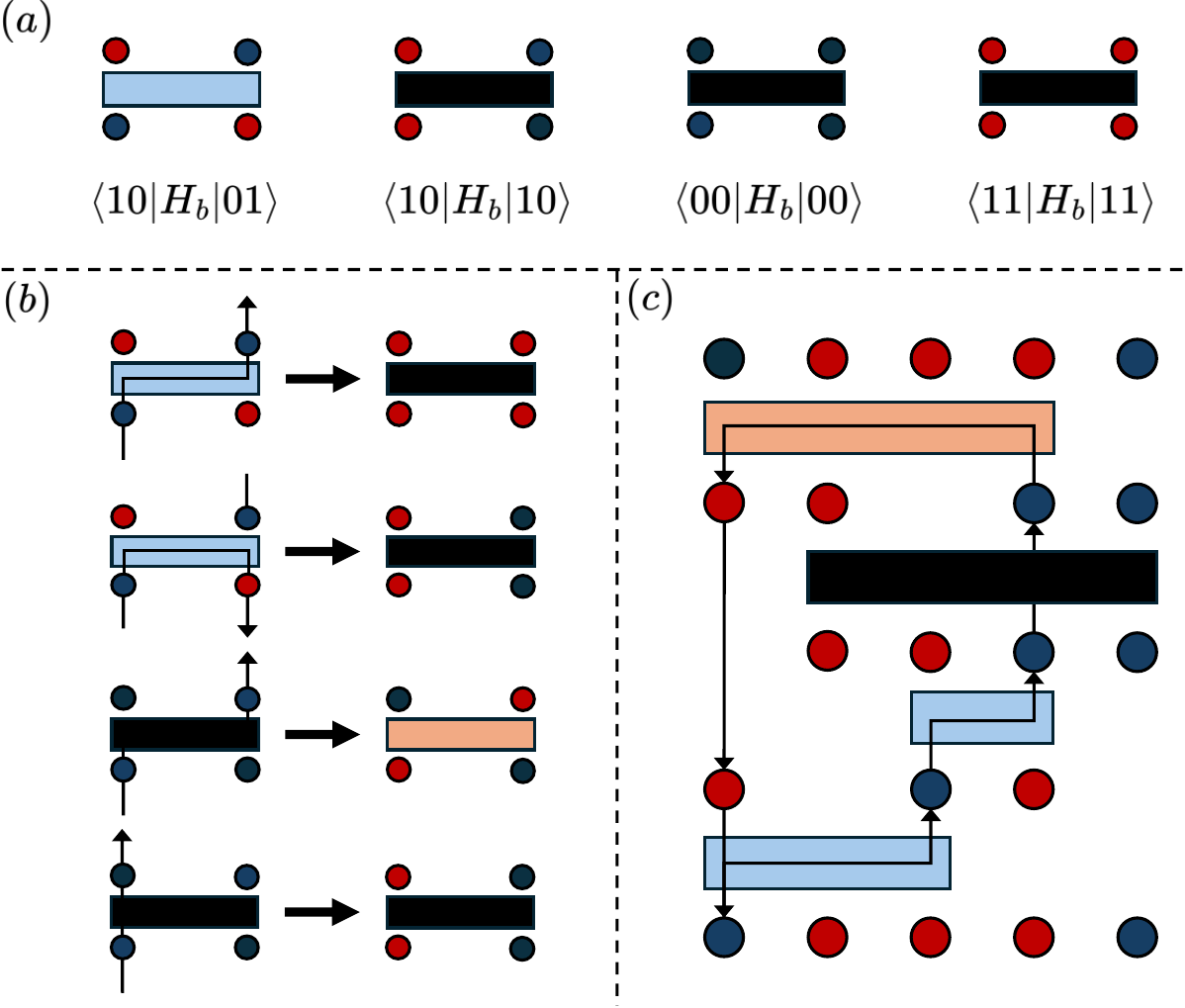}
    \caption{SSE directed loop construction for the XY model. Panel $(a)$ shows the dictionary of matrix elements and their corresponding diagrammatic representations. Each matrix element is associated with four legs. Blue dots denote $|0\rangle$ and red dots denote $|1\rangle$. The blue rectangles represent the matrix element $\langle 10|S_i^- S_{i+r}^+|01\rangle$ and the orange rectangles represent its Hermitian conjugate. Black rectangles correspond to diagonal matrix elements. All four matrix elements shown in panel $(a)$ as well as their Hermitian conjugates are allowed for the XY model. Panel $(b)$ shows the possible transitions for the first and third matrix elements in panel $(a)$. Panel $(c)$ finally exhibits one possible closed loop that can be constructed for the XY model in the off-diagonal SSE update.}
    \label{fig:SSE_directed_loop}
\end{figure}

In the SSE algorithm, there are two types of moves \cite{sandvik2002directedloops, sandvik1999OperatorLoop}. The first allows for variations in the expansion order $n$ by inserting and removing diagonal operators $H_{1,b}$. In our approach, we use a heatbath scheme instead of the Metropolis updates typically favored for nearest-neighbor models because power-law decay interactions can lead to larger rejection probabilities \cite{sandvik2019Review2}. In particular, the two-step heatbath method allows us to pre-compute and index the required probability tables instead of calculating probabilities on-the-fly \cite{ejaaz2024sse, merali2025quantum}.

\begin{figure*}
    \centering
\includegraphics[width=0.47\linewidth]{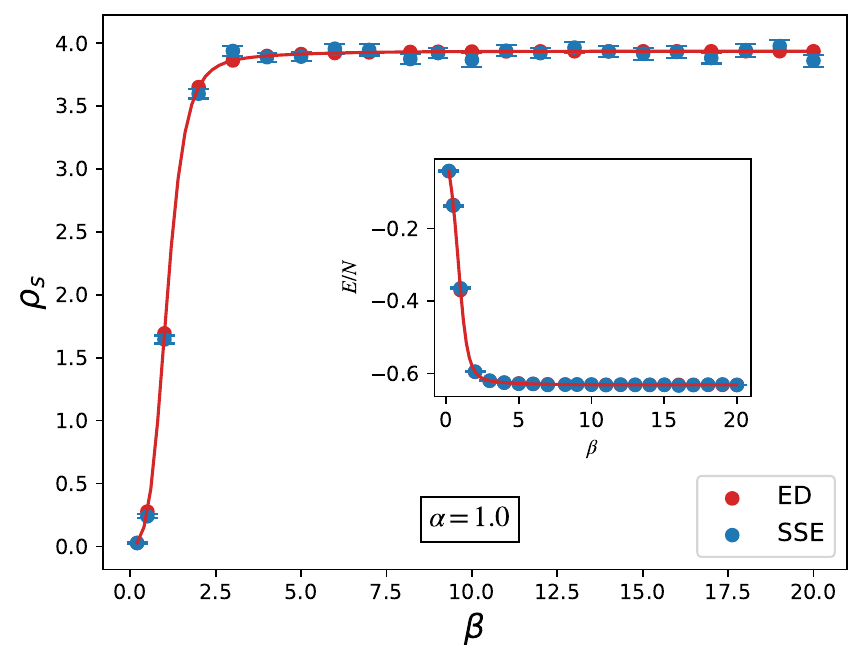}
\includegraphics[width=0.48\linewidth]{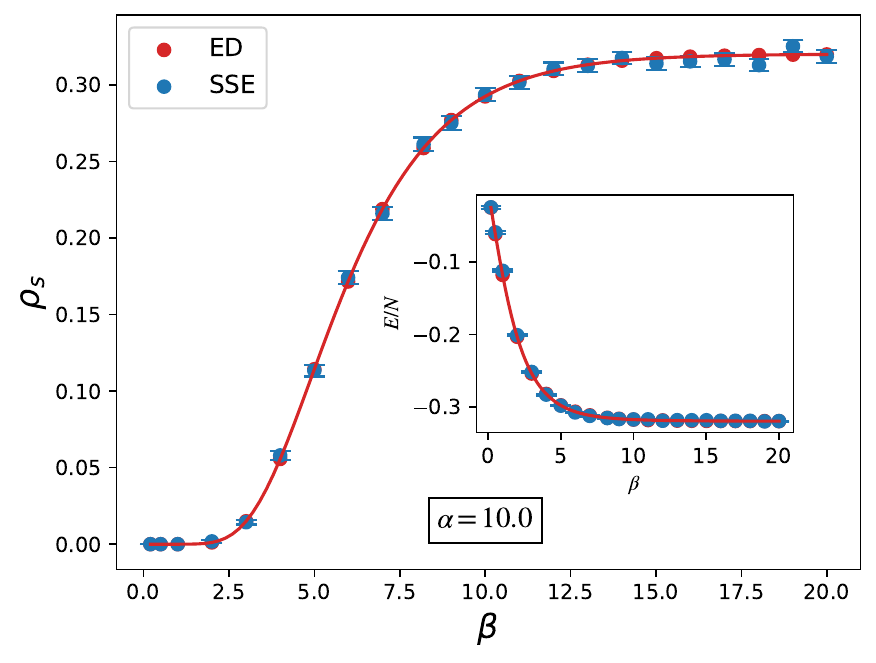}
\caption{QMC superfluid density and energy per site (insets) as a function of $\beta$ compared to ED for $N=11$. The left panel shows the results for $\alpha = 1$ and the right panel corresponds to $\alpha = 10$.} \label{fig:ED_QMC_stiffness}
\end{figure*}

The other type of move required for sampling the full configuration space is the off-diagonal update \cite{ melko2013SSEReview, sandvik1999OperatorLoop}. This proceeds by replacing the diagonal operators added during the first update with off-diagonal ones probabilistically. We accomplish this by leveraging the directed loop technique which has been shown to be particularly effective for systems with a $U(1)$ symmetry in past work \cite{sandvik2002directedloops, generalizeddirectedloops}. This update proceeds by associating with each $2$-body matrix element $\langle \sigma^i_k, \sigma^j_k|H_{b}|\sigma_{k+1}^i, \sigma_{k+1}^j\rangle$, a leg for each of the spin configurations $\sigma_r^s$ with $s \in \{i,j\}$ and $r \in \{k, k+1\}$ as shown in panel $(a)$ of Fig. \ref{fig:SSE_directed_loop}. 

A loop is constructed by entering at one of the four operator legs in the full SSE configuration and exiting at another one. In doing this, the entrance and exit legs are flipped: $ \sigma_{r}^s \rightarrow \sigma_{r}^s \oplus 1$ \cite{sandvik2002directedloops}. Panel $(b)$ of fig. \ref{fig:SSE_directed_loop} shows the possible transitions and exit legs for the first and third operator matrix elements in panel $(a)$, associated with the bottom left entrance leg. Given an entrance leg, the probability of exiting at any of the four possible legs is determined by the initial and final matrix elements to satisfy detailed balance. Then, the simulation proceeds by continuing to the next non-trivial matrix element: $k \rightarrow k+l$ and the next entrance leg corresponds to the same site as the exit leg from the previous operator \cite{sandvik1999OperatorLoop}.

Note that for the XY model, for a given bond, all matrix elements are identically $1/2r_b^\alpha$. Therefore, this off-diagonal update does not change the many-body configuration weight in our case. This yields two tremendous simplifications in the directed loop update scheme. The first is that bounces (a transition such that the exit leg is the same as the entrance leg) can be ignored \cite{sandvik1999OperatorLoop}. The second is that the entire $(1+1)d$ SSE configuration space can be sub-divided into the directed loops constructed as outlined above, and the resulting loops flipped independently with a probability of $1/2$ \cite{groundstate1999sandvik}. Even though the off-diagonal update does not change the configuration weight here (as mentioned above), it is instrumental for achieving an efficient traversal of the SSE sample space \cite{groundstate1999sandvik, generalizeddirectedloops, melko2013SSEReview}. 

One possible loop for the $1d$ XY model with $N=5$ is exhibited in panel $(c)$ of Fig. \ref{fig:SSE_directed_loop} for reference. The SSE configuration has four matrix elements in this case, three of which are off-diagonal and participate in the constructed loop. Note that diagonal matrix elements are not excluded from the operator loop construction in general. For each of the operators in the loop, there are two choices for the exit leg. One of these is selected in the construction with a probability of $1/2$. During the construction of this loop, the encountered legs are flagged. After this loop is constructed in practice, the simulation proceeds with the next un-flagged leg in the stored operator list until all the legs have been flagged. 

The energy estimator in SSE is closely related to the expansion order \cite{sandvik1992generalization}: 
\begin{equation}
    \langle E \rangle = -\frac{\langle n \rangle}{ \beta} + JC.
\end{equation}
Here, we have added the offset $JC$ to correct the energy estimate. Expectation values of diagonal estimators $O$ such as the magnetization can be evaluated simply using the spin configuration of the first time-slice: $\langle O \rangle = \langle O(\alpha) \rangle$ \cite{Sandvik2010Review}. Here, the left side denotes the operator expectation value and the right side corresponds to the statistical average over all simulation steps of the matrix elements $O(\alpha) = \langle \alpha |O | \alpha \rangle$. The superfluid density estimator is related to the winding number $W$ \cite{melko2004aspect, sandvik1997finite, sandvik1992generalization}:    
\begin{equation}
    \rho_s = \frac{1}{\beta N} \langle W^2 \rangle.
\end{equation}

In the XY model with arbitrary interactions, the winding number can be computed via a bond-distance weighted accumulation of the number of each type of off-diagonal operators. To this end, we define:
\begin{gather}
    N_W(b) = \begin{cases}
    r_b \ \ \text{if} \ H_b = S_{i}^{-} S_{j}^{+}, r_b \leq (N-1)/2 \\
    -r_b \ \ \text{if} \ H_b = S_{i}^{+} S_{j}^{-}, r_b \leq (N-1)/2 \\
    -r_b \ \ \text{if} \ H_b = S_{i}^{-} S_{j}^{+}, r_b > (N-1)/2 \\ 
    r_b \ \ \text{if} \ H_b = S_{i}^{+} S_{j}^{-}, r_b > (N-1)/2 \\ 
    0 \ \ \text{if} \ H_b = I.
    \end{cases}
\end{gather}
Here, $r_b = j(b) - i(b)$ with $j > i$ and $i,j \in \{1,...,N\}$. $N_W$ can be thought of as a distance counter for hardcore bosons hopping through the lattice. Note that the third and fourth cases of the above equation follow simply from Eq. \eqref{twisted_long_range} since the sum over $r$ in that Hamiltonian goes from $1$ to $(N-1)/2$. The terms corresponding to bonds $b = (i_1(b), i_2(b))$ with distance $i_2(b) - i_1(b) > (N-1)/2$ are counted then by the sum over $j$ in Eq. \eqref{twisted_long_range}, which gives a positive phase twist for the pair of sites $(i_2(b), i_1(b))$. This, in turn, corresponds to a negative phase twist on $b = (i_1(b), i_2(b))$. The winding number is then given by:
\begin{equation}
    W = \sum_{j=1}^{M} \sum_{b=1}^{N_b} N_W(b).
\end{equation}

To validate our QMC algorithm, we compare the energy per site and superfluid density estimates to ED for small system sizes for a range of temperatures ($\beta = J/T \in [0.2,20]$). The results for $\alpha = 1$ (left panel) and $\alpha=10$ (right panel) with $N=11$ are shown in Fig. \ref{fig:ED_QMC_stiffness}. For this entire set of temperatures, we see good convergence to ED results. Additionally, it is interesting to note that for both values of $\alpha$, for a sufficiently large temperature, the superfluid density converges to $0$. However, as $T \rightarrow 0$, $\rho_s(\alpha = 1) > \rho_s(\alpha = 10)$. We will examine this superfluid density enhancement in more detail in the thermodynamic limit in Sec. \ref{sec:stiffness_enhancement}. 

\section{Mean Field \& Spin Wave Analysis \label{sec:approx_sols}}

\subsection{\label{sec:mft} Mean Field Theory}
In the thermodynamic limit ($N \rightarrow \infty$), the uniformly interacting ($\alpha = 0$) case should converge to the MFT result \cite{goldenfeld2018lectures}. In this section, we examine the convergence of the exact LMG solution as $N \rightarrow \infty$. To determine the mean field behavior, consider again the Hamiltonian of the uniformly interacting model: 
\begin{gather}
    H_0 = -\frac{1}{2}\sum_{i \neq j} J_{ij} (S_i^x S_j^x + S_i^y S_j^y) \nonumber \\ = -\frac{1}{2} \sum_{i \neq j} J_{ij} \ \left(\vec{S_i} \cdot \vec{S_{j}} \right).
\end{gather}
Here, we have simply collected the in-plane spin components into vectors: $\vec{S_i} = \left(S_i^x, S_i^y\right)$. The mean field solution can be obtained by adding a field in the XY plane and ignoring fluctuations. Then, minimizing the free energy and solving the resulting equation in the limit of $T \rightarrow 0$, we get the following for the ground state energy and superfluid density (see Appendix \ref{sec:appendix_B_mft}): 
\begin{gather}
    E_0 = -\frac{JN (N-1)}{8}, \label{mft_energy} \\
    \rho_s = \frac{JN}{8} \left(\frac{N^2-1}{12} \right). \label{mft_rho_s}
\end{gather}

\begin{figure*}
    \centering
    \includegraphics[width=0.48\linewidth]{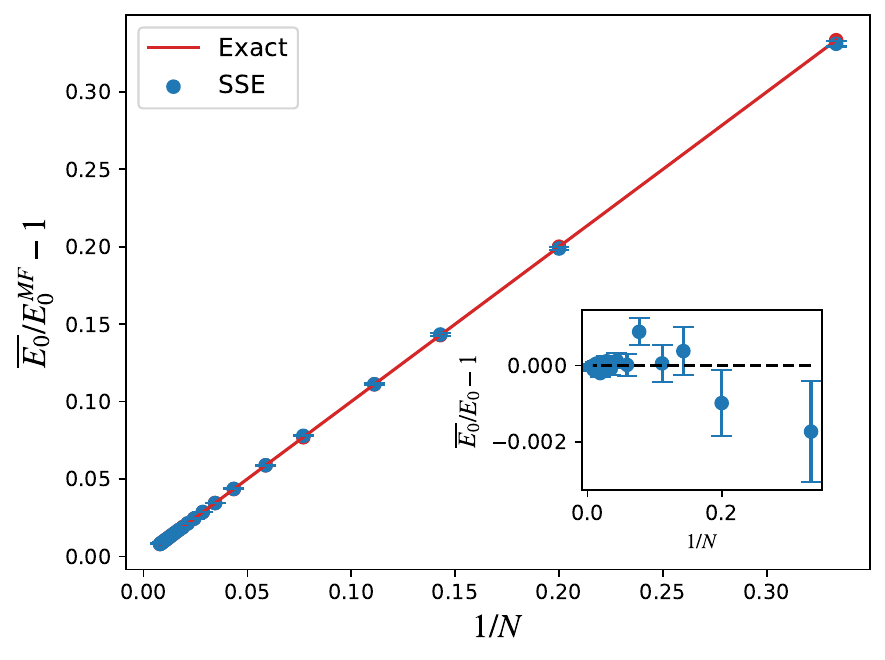}
    \includegraphics[width=0.48\linewidth]{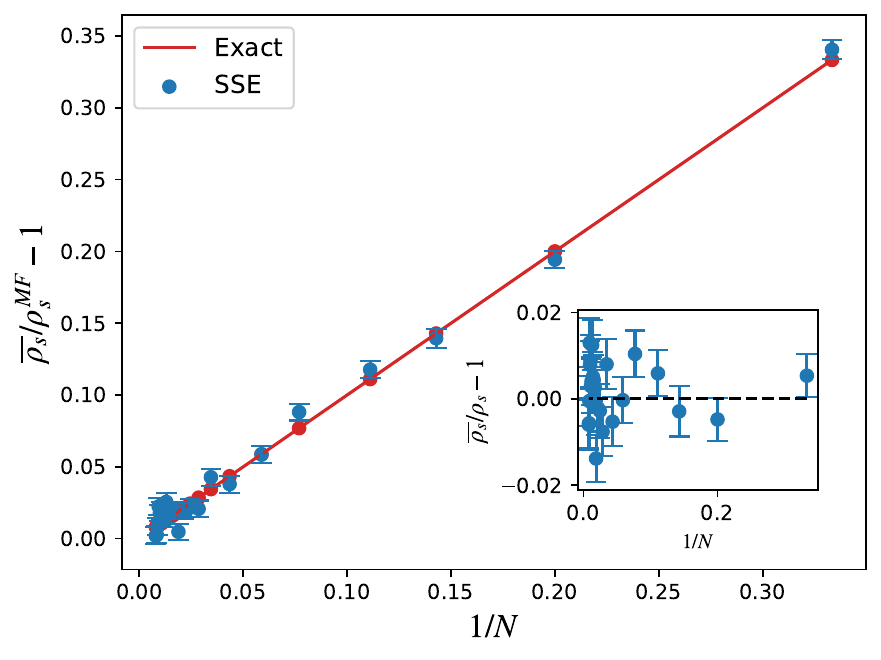}
    \caption{Convergence of the QMC estimates of ground state energy (left panel) and superfluid density (right panel) to the mean field results in the limit of $N \rightarrow \infty$. The QMC estimates of energy and superfluid density are denoted with a bar: $\overline{E_0}$ and $\overline{\rho_s}$, and the mean field results are denoted by $E_0^{MF}$ and $\rho_s^{MF}$. The exact finite size effect in $N$ (given by Eq. \eqref{finite_size_mft}) is also shown for comparison for both $E_0$ and $\rho_s$ by the red lines. The insets show the comparison between QMC estimates and LMG predictions. Here, the denominators $E_0$ and $\rho_s$ correspond to the (exact) LMG results.}
    \label{fig:mft_qmc_comparison}
\end{figure*}

We will refer to the above estimates as the MFT or classical results throughout the rest of this work. Comparing to the LMG model, we see that both $E_0$ and $\rho_s$ converge to the mean field result with an $O(1/N)$ finite size effect: 
\begin{equation}
    E_0/E_0^{MF} = \rho_s/\rho_s^{MF}= 1 + 1/N. \label{finite_size_mft}
\end{equation} 

It is instructive to use this finite size effect to test our QMC algorithm for $\alpha=0$. Fig. \ref{fig:mft_qmc_comparison} shows the convergence in $N$ for $\alpha=0$ to the MFT results since we expect our QMC results to converge to the LMG limit. The left panel shows the energy ratio $\overline{E_0}/E_0^{MF}$ where $\overline{E_0}$ represents the QMC estimates for each $N$ with $\beta = 20$. The exact scaling of this ratio in $N$ from Eq. \eqref{finite_size_mft} is also exhibited for reference (denoted by the red line). The inset shows the ratio $\overline{E_0}/E_0$ where $E_0$ is the exact LMG result given by Eq. \eqref{energy_gs_lmg}, and discussed in Sec. \ref{sec:lmg}. 

Similarly, the right panel shows the superfluid density estimate ratio $\overline{\rho_s}/\rho_s^{MF}$ which has the same scaling as the energy, given by Eq. \eqref{finite_size_mft}. The QMC results are again denoted by $\overline{\rho}$ and the red line exhibits the finite size effect from Eq. \eqref{finite_size_mft}. The inset shows the QMC estimates of the superfluid density divided by the exact LMG result from Eq. \eqref{lmg_stiffness}. In all of these cases, we see good convergence to the theoretically predicted results. The statistical error bars for the ratio $\rho_s/\rho_s^{MF}$ and $\rho_s/\rho_s^{LMG}$ are larger than the energy since it converges slower in simulation time. Moreover, note that our calculation for the superfluid density is perturbative while our finite size effect calculation for the energy is exact. Therefore, higher order effects also contribute to the deviations observed in the superfluid density ratio. Despite the fluctuations, it is clear that the QMC results converge to the MFT results as $N \rightarrow \infty$. 

\subsection{\label{sec:lsw_approx} Linear Spin Wave Theory}

We can extend our analysis beyond the classical mean field solution to include quantum fluctuations using LSW \cite{coletta2012semiclassical, bernardet2002analytical, TommasoLSWLRXY, Santos1987lswXY}. To this end, consider again the power-law decay XY model Hamiltonian in the presence of a twist: 
\begin{equation}
    H(\theta) = -\frac{J}{2} \sum_{i=0}^{N-1} \sum_{r \neq 0} \frac{\cos(r \theta)}{|r|^\alpha} (S_x^i S_x^{i+r} + S_{y}^{i} S_{y}^{i+r}). \label{lsw_Hamiltonian_1}
\end{equation}
Here, we have symmetrized the sum from Eq. \eqref{sum_pairs_H}, and defined $j = i+r$ with $r \in [-(N-1)/2, (N-1)/2]$, $r \neq 0$ for the periodic chain with $N$ odd. We have also neglected the current term associated with a non-zero twist in the Hamiltonian since it does not contribute to the ground state energy, as before. To construct the LSW theory, we proceed by mapping our spin operators to Holstein-Primakoff bosons \cite{HolsteinPrimakoffBosons}. The resulting Hamiltonian can then be diagonalized using a Bogoliubov transformation after neglecting higher order terms (see Appendix \ref{sec:appendix_C_lsw_diagonalization}). 

The ground state energy is: 
\begin{equation}
    E_0 = -\frac{JN \gamma_0}{4} + \sum_k \left(\sqrt{A_k^2 - B_k^2} - A_k\right). \label{lsw_energy}
\end{equation}
Here, $A_k$, $B_k$ and $\gamma_k$ are given by: 
\begin{gather}
    A_k(\theta) = \frac{J}{2} \left(\gamma_0(\theta) - \frac{\gamma_k(\theta)}{2} \right), \ \ B_k = \frac{J \gamma_k(\theta)}{4}, \\ \gamma_k(\theta) = \sum_{r=1}^{(N-1)/2} \frac{\cos(2 \pi r k/N) \cos(r \theta)}{r^\alpha}.  \label{gamma_k_full_def}
\end{gather}

It is interesting to analyze the case for $\alpha=0$ with LSW theory. The first term of the ground state energy in Eq. \eqref{lsw_energy} recovers exactly the mean field result from Eq. \eqref{mft_energy} as expected. Adding in the LSW correction term yields the following estimate for the ground state energy (see Appendix \ref{sec:appendix_C_lsw_alpha_0} for details): 
\begin{gather}
    E_0^{LSW} = -\frac{J(N^2-1)}{8}. \label{lsw_e0_alpha_0}
\end{gather}
This coincides exactly with the LMG ground state energy given by Eq. \eqref{energy_gs_lmg}. To analyze the superfluid density associated with the $\alpha=0$ limit using LSW theory, we need to evaluate the ground state energy in the presence of a small twist $\theta$. As before, evaluating the superfluid density using finite differences gives (see Appendix \ref{sec:appendix_C_lsw_alpha_0} for details): 
\begin{equation}
    \rho_s^{LSW} = \frac{J(N+1)}{8} \left (\frac{N^2 - 1}{12} \right), \label{lsw_rho_alpha_0}
\end{equation}
as expected from the LMG limit analysis in Sec. \ref{sec:lmg}. It is instructive to note that simply using the first term in Eq. \eqref{lsw_energy} to evaluate the LSW energies yields the MFT/classical result for the superfluid density. Comparing the LSW theory estimates for $\alpha=0$ (Eqs. \eqref{lsw_e0_alpha_0} and \eqref{lsw_rho_alpha_0}) to the LMG limit (Eqs. \eqref{energy_gs_lmg} and \eqref{lmg_stiffness}), it is clear that LSW theory gives the exact energy as a function of $N$. Consequently, the LSW correction reproduces the $1/N$ finite size correction derived by comparing the LMG limit and mean field results. 

\begin{figure}
    \centering
    \includegraphics[width=\linewidth]{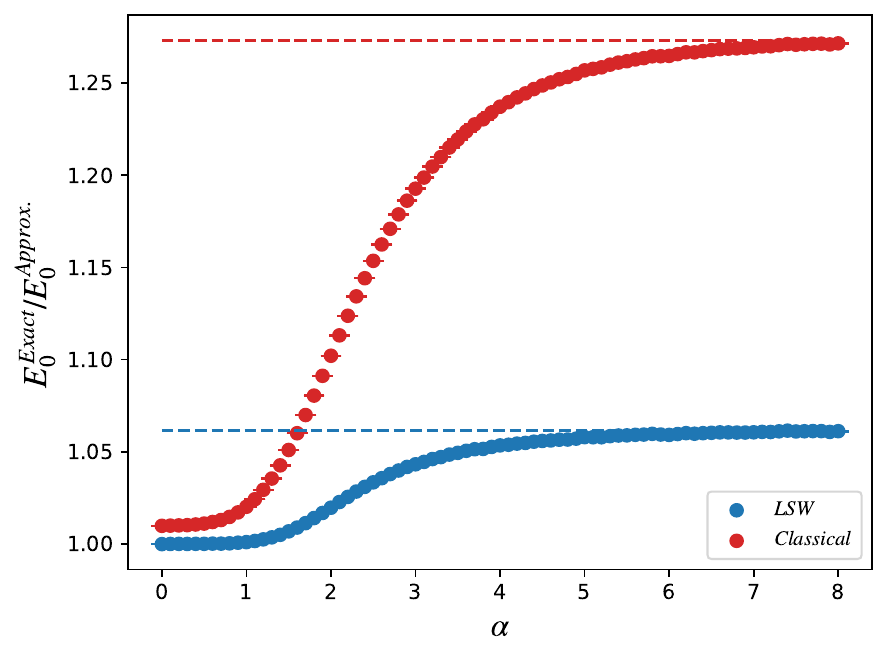}
    \caption{$E_0^{\text{Exact}}/E_0^{\text{Approx.}}$ as a function of $\alpha$ for $N=101$. The exact results are computed using QMC with $\beta = 200$. The approximations here correspond to the classical solution and the LSW corrected estimates. The lines correspond to the limit $\alpha \rightarrow \infty$ for which the exact results are computed using the FFS. The red line is associated with the classical solution and the blue line corresponds to the LSW theory estimate.}
    \label{fig:lsw_QMC_comparison}
\end{figure}

It is also interesting to examine the deviation of LSW theory from the exact solution (Sec. \ref{sec:fermion_sol}) for the nearest-neighbor model. In that limit, $\gamma_0 \rightarrow 0$ and $\gamma_k \rightarrow \cos(2 \pi k/N)$ \cite{coletta2012semiclassical, bernardet2002analytical}. Since the twist for a nearest-neighbor model is no longer bond dependent, it can be trivially absorbed into the coupling constant $J \rightarrow J\cos(\theta)$. Expanding $\cos(\theta) \approx 1 - \theta^2/2$, we get $\rho_s = -E_0/N$. Setting $J=1$, the classical solution corresponding to $\alpha \rightarrow \infty$ gives $\rho_s = 0.25$ while the exact FFS (Sec. \ref{sec:fermion_sol}) yields $\rho_s = 1/\pi \approx 0.3183$. Therefore, the classical solution has an error of $\sim 21.5 \%$. Adding in the LSW correction (computed numerically using $N \sim 10^4$ sites) gives $\rho_s = 0.2998$ which reduces the error down to $\sim ~ 5.84 \%$. The error in the nearest-neighbor limit is fairly large since we are considering a $1d$ lattice where higher order fluctuations need to be accounted for exactly for quantitative accuracy.

To examine the LSW improvement over the classical solution quantitatively across a wide range of $\alpha$, we consider the ratio $E_0^{\text{Exact}}/E_0^{\text{Approx.}}$. Fig. \ref{fig:lsw_QMC_comparison} exhibits this for $\alpha \in [0.0, 8.0]$ with $N=101$. The exact values correspond to QMC results with $\beta = 200$. The asymptotes for each of the curves in the limit of $\alpha \rightarrow \infty$ computed using nearest-neighbor estimates discussed above are also shown for comparison. The LSW corrections tremendously improve the accuracy of the approximation for the entire range of $\alpha$ exhibited. Moreover, the error increases with $\alpha$ for both the classical solution as well as the LSW corrected estimate, saturating in the nearest-neighbor limit. 

This is unsurprising since for $\alpha = 0$, the system converges to the mean field limit for large $N$ which coincides with the classical solution. Here, the LSW correction simply provides the finite size correction. In the opposite limit, $\alpha \rightarrow \infty$, the system converges to the nearest-neighbor case where higher order effects are more pronounced. This increasing error with $\alpha$ for LSW theory in the regime of interest for trapped-ion experiments ($1 \leq \alpha \leq 3$) is what necessitates the development of computational approaches such as QMC for quantitative accuracy. 

\section{\label{sec:stiffness_enhancement} Behavior of the Superfluidity Density}

\begin{figure}
\centering
\includegraphics[width=\linewidth]{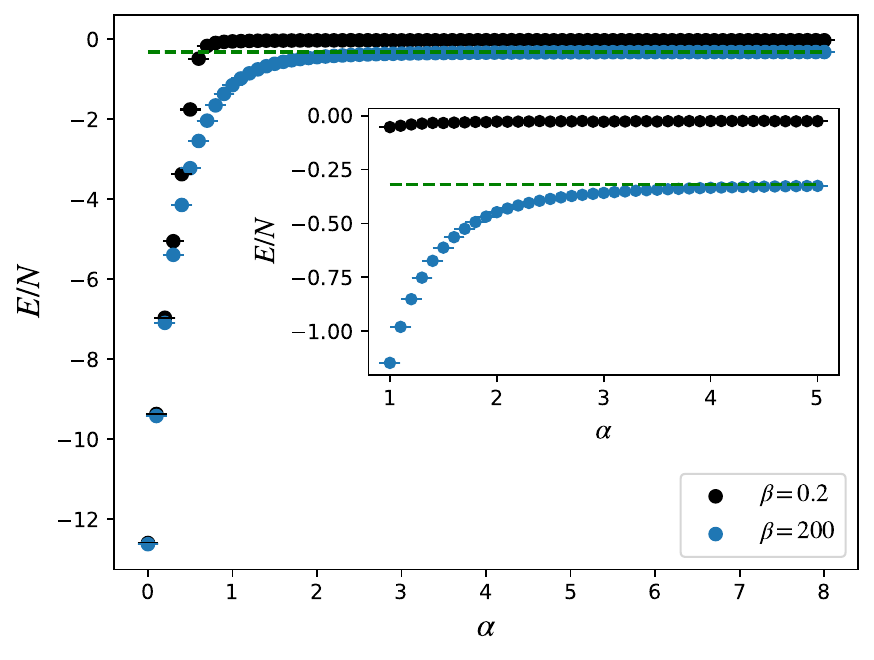}
\caption{QMC Energy per site estimates as a function of $\alpha$ for $N=101$, $\alpha \in [0,8]$. The inset magnifies the results for $\alpha \in [1,5]$. Curves for two different values of $\beta$ are exhibited. For $\beta = 200$, the energy converges to the free-fermion solution as $\alpha \rightarrow \infty$, indicated by the dashed line. For $\beta = 0.2$, the energy is larger due to the higher temperature, as expected. For all of these simulations, $0.5 \times 10^6$ equilibration steps and $0.5 \times 10^6$ MC steps were used.}
\label{fig:QMC_Energy}
\end{figure}

To study the behavior of the superfluid density in the phase diagram as a function of $\alpha$, we perform large scale QMC simulations with $N=101$. Fig. \ref{fig:QMC_Energy} shows the ground state energy per site for $\beta \in \{0.2, 200\}$. 
The inset magnifies the energy per site for $\alpha \in [1.0, 5.0]$. The ground state corresponds to the $M_z = 0$ sector, which we have confirmed by observing magnetization data. 

The energy per site for $\beta = 200$ converges exactly to the ground state energy predicted by the free-fermion solution in the $\alpha \rightarrow \infty$ limit (exhibited by the dashed line in Fig. \ref{fig:QMC_Energy}). Note that convergence in temperature to the ground state for a given $N$ is slower when $\alpha$ is larger, as shown in Fig. \ref{fig:ED_QMC_stiffness}. Therefore, since our ground state energy is converged for $\alpha = 8.0$ when $\beta = 200$, we can safely surmise that the QMC energy estimates for $\alpha < 8.0$ reflect the ground state energy per site. Moreover, we have also confirmed that the energy per site for $\beta = 20$ agrees within statistical errors with the curve for $\beta = 200$, further indicating that $\beta = 200$ is sufficient for energy convergence to the ground state. 

For $\beta = 0.2$, the energy per site is higher than the ground state energy for $\alpha > 1$, as expected. In the limit of $\alpha \rightarrow 0$, the energy per site for $\beta = 0.2$ converges to the ground state energy as well. This can be understood by considering the gap between the ground state and the excited states. Using LSW, for $\alpha = 0$, we get that the excitation energy for any mode $k$, denoted $\epsilon_k$, scales linearly with $N$ (see Eq. \eqref{epsilon_k_alpha_0} in Appendix \ref{sec:appendix_C_lsw_alpha_0}). Expectation values in QMC are calculated using the canonical ensemble: 
\begin{equation}
    \langle O \rangle = \frac{\langle O \rangle_0 + \sum_{i=1}^{2^{N}-1} e^{-\beta \Delta E_i} \langle O \rangle_i}{1 + \sum_{i=1}^{2^{N}-1} e^{-\beta \Delta E_i}} . \label{canonical_ensemble_average}
\end{equation}

Here, we have defined $\langle O \rangle_i = \langle E_i |O| E_i \rangle$ for $0 \leq i \leq 2^{N}-1$. As $\alpha \rightarrow 0$, the gap $\Delta E_i \sim \mathcal{O}(N)$. Therefore, for both values of $\beta$ considered, $e^{-\beta \Delta E_i} \approx 0$ and we get the ground state energy. In contrast, for larger values of $\alpha$, the gaps are smaller \cite{TommasoLSWLRXY}. Consequently, the energy estimates for $\beta = 0.2$ are markedly different from those corresponding to $\beta = 200$. Finally, because of the power-law interactions, the energy of the system is super-extensive for $\alpha \leq 1$. This can be seen, for instance, by considering the first term (classical estimate) of the ground state energy from Eq. \eqref{lsw_energy}. For $\alpha \leq 1$, the series $\gamma_0 = \sum_{r=1}^{(N-1)/2} \frac{1}{r^{\alpha}}$ diverges in the limit of $N \rightarrow \infty$ which causes the energy per site to diverge as $\alpha \rightarrow 0$. 

\begin{figure}
    \centering
    \includegraphics[width=\linewidth]{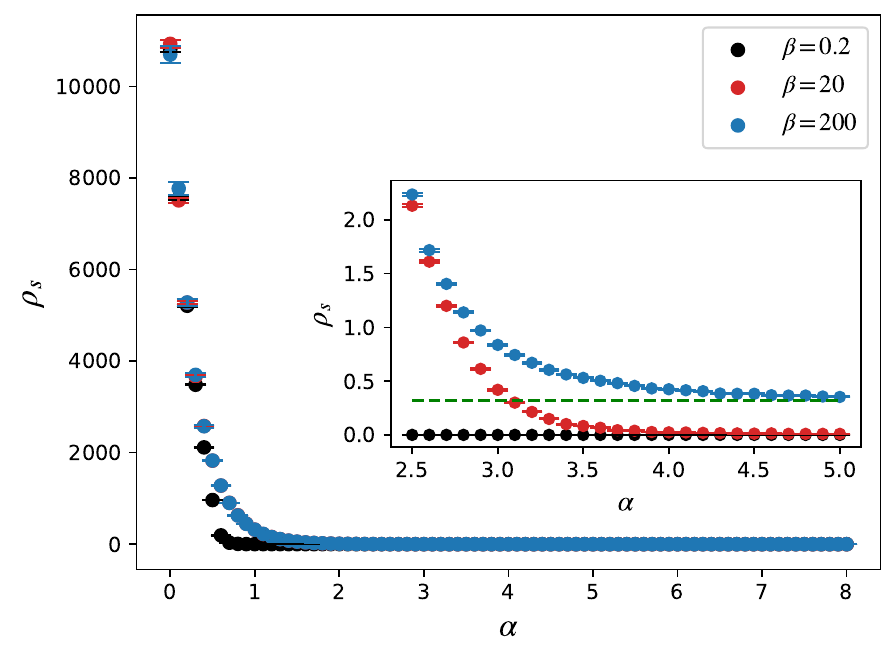}
    \caption{Superfluid density as a function of $\alpha$ for $N=101$ and $\beta \in \{0.2, 20, 200\}$ computed using QMC. The inset magnifies the curves corresponding to $\alpha \in [2.5, 5.0]$. For all of these simulations, $0.5 \times 10^6$ equilibration steps and $0.5 \times 10^6$ MC steps were used. The dashed line indicates the free-fermion solution estimate of the superfluid density.}
    \label{fig:QMC_stiffness_alpha}
\end{figure}

\begin{figure*}
    \centering
    \includegraphics[width=0.50\linewidth]{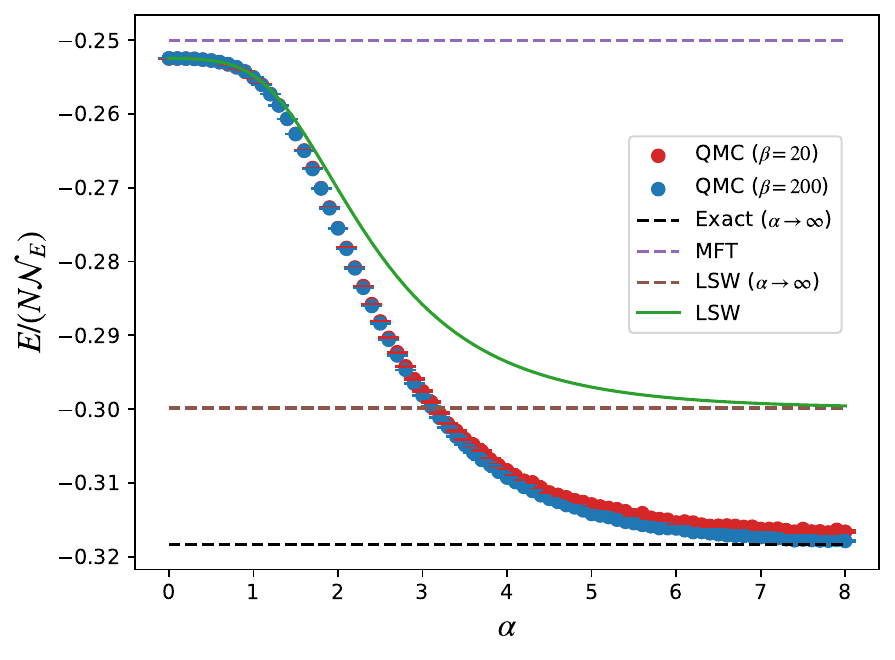}
    \includegraphics[width=0.48\linewidth]{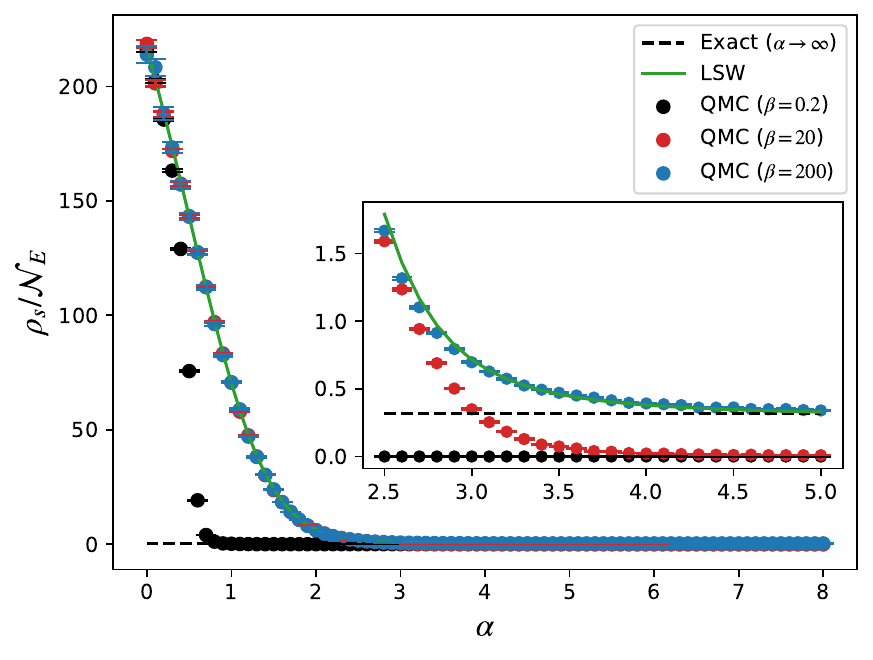}
    \caption{The left panel shows the Kac normalized ground state energy per site and the right panel shows the Kac normalized superfluid density. The QMC estimates were computed using $N=101$ for $\beta = 0.2$, $\beta = 20$ and  $\beta = 200$ as a function of $\alpha$. The exact normalized energy per site and the normalized superfluid density in the nearest-neighbor limit is shown using the black dotted line. The normalized MFT estimate of the ground state energy per site is shown using the purple dotted line. The LSW estimate is exhibited using the green solid line for both cases. The brown line shows LSW estimate of the energy per site in the limit of $\alpha \rightarrow \infty$.}
    \label{fig:kac_normalized_energy_stiffness}
\end{figure*}

Fig. \ref{fig:QMC_stiffness_alpha} shows the QMC estimates of $\rho_s$ as a function of $\alpha$. The inset magnifies the curve for $\alpha \in [2.5, 5.0]$ and the dashed green line again shows the free-fermion result. Here, we see that $\beta = 20$ is not sufficient to converge to the ground state superfluid density when $\alpha$ is large. However, the curve corresponding to $\beta = 200$ converges to the expected result. Moreover, for $\beta = 200$, a clear enhancement of the superfluid density can be observed for all values of $\alpha$ relevant for trapped ion implementations of the XY model ($1 \leq \alpha \leq 3$) compared to the nearest-neighbor limit described by FFS. In the QMC calculations, this presents as a diverging superfluid density for $\alpha < 3$ in the thermodynamic limit. 

This divergence itself can be understood by considering the second derivative of the zero mode $\gamma_0(\theta)$ in our LSW analysis. For the superfluid response, this is differentiated twice which yields the series $\gamma_0''(0) = \sum_{r=1}^{(N-1)/2} 1/r^{\alpha - 2}$ which clearly diverges for $\alpha \leq 3$. Recent work in LL theory also reflects this diverging superfluid response in the long-range model \cite{dupuis_2024}. For power-law interactions, the Luttinger parameter and velocity both acquire a momentum dependence: $K_{LL}, v \rightarrow K_{LL}(k), v(k)$, and diverge in the small $k$ limit as $\sim |k|^{(\alpha - 3)/2}$. This causes the superfluid density $\rho_s(k \rightarrow 0) = K_{LL}(k) v(k)/\pi$ to diverge as well suggesting an enhancement of superfluidity due to the long range hopping of bosons for $\alpha < 3$. 

It is also instructive to examine the behavior of the superfluid density curves for different temperatures in more detail. For $\beta = 0.2$, we see a trivial superfluid density for $\alpha > 1$. This indicates that the $\beta = 0.2$ curve corresponds to a temperature above the critical temperature $T_c$ for this system size ($N=101$), as discussed in Sec. \ref{sec:stiffness}. For $\beta = 0.2$ and $\alpha < 1.0$, our QMC results again converge to the ground state superfluid density because of the large energy gap in that regime. For $\beta = 20$ and $\beta = 200$, both curves are virtually identical for small $\alpha$. However, for larger $\alpha$, the two curves behave differently from each other; $\rho_s(\beta = 200)$ converges in $\alpha$ to the expected ground state results while $\rho_s(\beta = 20)$ converges to zero as $\alpha$ grows larger and the system converges to the nearest-neighbor limit. 

This is a manifestation of the competition between the limits $T \rightarrow 0$ and $N \rightarrow \infty$ (discussed in Sec. \ref{sec:stiffness}), and can be understood by appealing to the gapless nature of the ground state in the limit of $N \rightarrow \infty$ \cite{TommasoLSWLRXY}. As mentioned above (see Eq. \eqref{canonical_ensemble_average}), convergence to the ground state in QMC depends on $ \beta \Delta E_i$ for all energy eigenstates $E_i$. As $N$ grows larger, the gaps between low-lying excitations and the ground state decrease, necessitating an increase in $\beta$ to recover the same level of convergence to the ground state. Therefore, even though $\beta = 20$ is sufficient for $N=11$ (as can be seen from Fig. \ref{fig:ED_QMC_stiffness}), for $N=101$, we need to use a much larger value of $\beta$ to extract the ground state superfluid density. In particular, our results indicate that $\beta = 200$ is sufficient for convergence to the ground state superfluid density for $N = 101$. 

As a function of $\alpha$, the dispersion relation (from LSW theory) behaves as: $\omega(\alpha) \sim k^z \sim \mathcal{O}(1/N^z)$. For $\alpha < 1$, $z = 0$ and for $\alpha > 3$, $z = 1$ \cite{TommasoLSWLRXY}. In the intermediate range ($1 < \alpha < \alpha_c$), $z = (\alpha - 1)/2$ \cite{maghrebi2017continuous, TommasoLSWLRXY}. This modulation of the dispersion relation determines the critical temperature above which our superfluid density estimator (spin stiffness) converges to zero. For $\alpha < 1$, we observe that the superfluid density for $\beta = 0.2$ converges to the ground state result (the $\beta = 200$ curve) because the system becomes gapped. As $\alpha$ increases, $z$ increases and the system becomes gapless. This forces the superfluid density for $\beta = 0.2$ to converge to zero for $\alpha > 1$. Finally, as $z$ saturates to $1$ above $\alpha = 3$, it is clear that the critical temperature for $N = 101$ drops further as the rate at which the gap closes increases, and only the $\beta = 200$ curve converges to the ground state result. 

It is instructive to note that energy calculations are less sensitive to this effect. To see this, consider the QMC energy estimate: 
\begin{equation}
    \langle H \rangle \approx \frac{E_0 + \sum_{j=1}^{K-1} e^{-\beta \Delta E_j}E_j}{1 + \sum_{j=1}^{K-1} e^{-\beta \Delta E_j}}.
\end{equation}
Here, $K < 2^{N} - 1$ is the number the states for which the weights $e^{-\beta \Delta E_j}$ significantly contribute to the QMC estimate, given specific values of $\beta$ and $N$. As $\Delta E_j \rightarrow 0$, $e^{-\beta \Delta E_j} \rightarrow 1$ and $E_j \rightarrow E_0$ which yields: $\langle H \rangle \rightarrow E_0$. Therefore, since the averaging function here is the energy itself, the closing gap does not affect the ground state energy calculations in QMC.  

We know from the LMG limit analysis (Sec. \ref{sec:lmg}) as well as MFT (Sec. \ref{sec:mft}) that the ground state energy per particle is extensive for $\alpha \leq 1$ because $\gamma_0$ diverges in that domain. This is a consequence of the long-range interactions which cause the energy to become super-extensive. To remedy this, the Hamiltonian spectrum is often normalized using the Kac scheme: $E \rightarrow \tilde{E} = E/\mathcal{N}_E$ \cite{schuckert2025observation, ComparinRobust, LR_Defenu}. The normalization factor $\mathcal{N}_E$ forces the energy to be extensive for all $\alpha$ and is defined as follows \cite{kac1963van}: 
\begin{equation}
    \mathcal{N}_E = \frac{1}{N} \sum_{i < j} \frac{J_{ij}}{J} = \sum_{r=1}^{(N-1)/2} \frac{1}{r^\alpha} .
\end{equation}
Here, $J_{ij}$ is the interaction strength between sites $i$ and $j$, and $\mathcal{N}_E$ can be interpreted as the average interaction strength across the entire periodic chain \cite{schuckert2025observation}. Note that $\mathcal{N}_E = \gamma_0$ with $\theta = 0$ from Eq. \eqref{gamma_k_full_def}. As $\alpha \rightarrow \infty$, $\mathcal{N}_E \rightarrow 1$ which allows us to recover the correct nearest-neighbor energy estimate. Moreover, in the limit of $\alpha \rightarrow 0$, $\mathcal{N}_E \rightarrow (N-1)/2$ which again restores the extensivity of the mean field energy estimate from Eq. \eqref{mft_energy} as desired. 

The left panel of Fig. \ref{fig:kac_normalized_energy_stiffness} shows the normalized ground state energy per site as a function of $\alpha$. We also exhibit the mean field normalized energy per site and nearest-neighbor ground energy per site limits. The green line shows the LSW estimate of the renormalized energy. As $\alpha \rightarrow 0$, the QMC estimates converge to the MFT results and as $\alpha \rightarrow \infty$, they converge to the free-fermion limit (where the normalization is trivial). LSW estimates overestimate the energy per site for all $\alpha$ as before and converge to the correct value as $\alpha$ decreases. 

The right panel shows the normalized superfluid density. We also show the LSW estimates of the superfluid density (computed using finite $N=101$) here which closely agree with the QMC estimates. While the normalization forces the energy to become extensive, the superfluid density remains super-extensive and diverges for $\alpha < 3$. Consequently, in the normalized version of the model ($J \rightarrow J/\mathcal{N}_E$) as well, we see an enhancement of superfluidity as $\alpha$ is decreased.  

In both the superfluid density results discussed above (Fig. \ref{fig:QMC_stiffness_alpha} and Fig. \ref{fig:kac_normalized_energy_stiffness}), the behavior of the superfluid response is dominated by the divergence for $\alpha < 3$. Therefore, it is instructive to examine the finite contributions to the superfluid density as a function of $\alpha$ instead. We can achieve this by introducing a normalization factor that cancels the diverging contribution $\gamma_0''(0) = \sum_{r=1}^{(N-1)/2} 1/r^{\alpha - 2}$, and forces the resulting quantity to be intensive as in the nearest-neighbor case. Note that this is similar to the Kac normalization discussed above to restore the extensivity of the energy. To this end, we define the normalized superfluid density as: $\tilde{\rho_s} = \rho_s/\mathcal{N_\rho}$. The normalization factor $\mathcal{N}_\rho$ is defined as: 
\begin{equation}
    \mathcal{N}_\rho = J\gamma_0''(0) = J\sum_{r=1}^{(N-1)/2} \frac{1}{r^{\alpha - 2}} .
\end{equation}

In the nearest-neighbor case ($\alpha \rightarrow \infty$), $\mathcal{N}_\rho \rightarrow J$ as required. This normalization removes the $\alpha$ dependent diverging factor $\gamma_0''(0)$ just as the Kac normalization removes the diverging contribution $\gamma_0(0)$ in the ground state energy. It is instructive to note that the superfluid density $\rho_s$ is proportional to the energy scale $J$ because it is the second derivative of the ground state energy. The normalization $\mathcal{N}_\rho$ also removes this dependence on $J$. In particular, for the nearest-neighbor limit, $\tilde{\rho_s} \rightarrow \rho_s(J, \alpha \rightarrow \infty)/J = \rho_s(J=1, \alpha \rightarrow \infty) = 1/\pi$. Consequently, since $\tilde{\rho_s}$ does not depend on $J$, it is also unaffected by the Kac normalization ($J \rightarrow J/\mathcal{N}_E$) discussed above, and yields the same result regardless of whether the Kac normalization scheme has been used to set the energy scale $J$.

\begin{figure}
    \centering
    \includegraphics[width=1.0\linewidth]{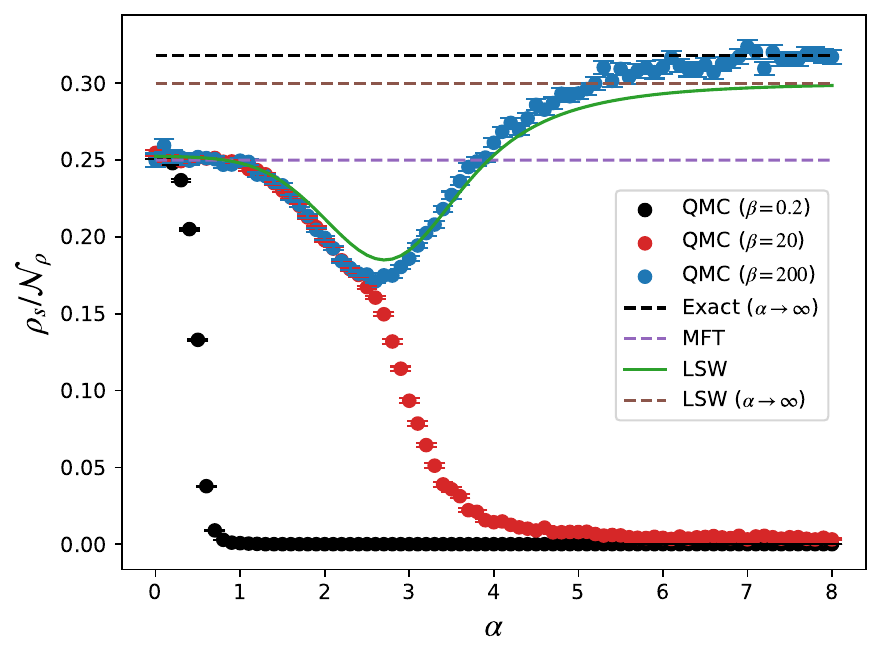}
    \caption{Superfluid density $\rho_s/\mathcal{N_\rho}$ normalized by the diverging factor $\gamma_0''(0)$ as function of $\alpha$. The QMC results correspond to $N=101$ with $\beta = 0.2$, $\beta = 20$ and $\beta = 200$. The black dotted line shows the nearest-neighbor superfluid density computed using the FFS. MFT and LSW estimates are shown using the purple dotted and green solid lines respectively. The brown line shows the LSW estimate in the limit $\alpha \rightarrow \infty$.}
    \label{fig:normalized_rho}
\end{figure}

Fig. \ref{fig:normalized_rho} shows this normalized response as a function of $\alpha$ for $N=101$. We also exhibit the MFT theory result after normalization which is constant as discussed above. The black dotted line exhibits the nearest-neighbor limit exact solution in the $\alpha \rightarrow \infty$ limit (computed using the FFS). The brown line shows the LSW theory estimate in the limit of $\alpha \rightarrow \infty$. As before, at $\beta = 0.2$, $T > T_c$ which causes the normalized superfluid response to be identically zero for $\alpha > 1$. For $\alpha < 1$, the results corresponding to $\beta = 0.2$ converge to the ground state normalized superfluid density results, as before. The $\beta = 20$ results converges to $0$ as $\alpha \rightarrow \infty$ which implies that $\beta = 20$ is not large enough to converge to the ground state when $\alpha > 3$. However, the curve corresponding to $\beta = 200$ yields the correct nearest-neighbor limit which implies good convergence to the ground state across the full range of $\alpha \in [0.0, 8.0]$. The ground state normalized response agrees with the MFT estimate for $\alpha \in [0,1]$ and with the nearest-neighbor estimate for $\alpha \rightarrow \infty$, as expected. 

The normalized superfluid density clearly distinguishes among the three regimes expected from LSW theory \cite{TommasoLSWLRXY}. The first is the long-range regime ($\alpha \leq 1$) where the normalized superfluid density is constant. The second is the medium range-regime ($1 < \alpha < \alpha_c$) where the normalized superfluid density monotonically decreases and acquires a minimum at $\alpha_c \approx 2.7$. The last regime corresponds to the short range model ($\alpha > \alpha_c$) which exhibits the same thermodynamic behavior as the nearest-neighbor XY model. Here, the normalized superfluid density monotonically increases until it converges to the superfluid density expected from the FFS. In the long-range regime, the normalized response $\tilde{\rho}$ is identically $0.25$ (the MFT result). It is also instructive to note that our transition point $\alpha_c \approx 2.7$ is less than $3.0$ as expected from the RG analysis of Ref. \onlinecite{maghrebi2017continuous}. Finally, we also exhibit the LSW theory estimate in Fig. \ref{fig:normalized_rho} which agrees with our QMC result qualitatively. However, it overestimates the normalized superfluid density in the medium range model and underestimates the normalized as well as the (un-normalized) superfluid density in the short-range model.

\section{\label{sec:conclusion} Conclusions}

In this work, we studied the power-law decay XY model focusing on the ground state superfluid density. The nearest neighbor variant exhibits a non-zero superfluid density as can be shown using the Jordan-Wigner transform \cite{jordan1928paulische, laflorencie2004scaling}. As the interaction strength exponent $\alpha$ is decreased, the range of interactions increases and this yields an enhancement of the superfluid density. We also analyzed the uniformly interaction limit ($\alpha = 0$) which yields the LMG model \cite{LMG2008}. The resulting Hamiltonian corresponds to a collective spin degree of freedom spanning the entire chain and can be solved exactly. The response to a uniform twist in this limit can then be determined perturbatively since the superfluid density is a linear response property. We found that the superfluid density is super-extensive in this limit (scales as $\mathcal{O}(N^3)$) and that the MFT results converge to the LMG superfluid density with an $\mathcal{O}(1/N)$ scaling. 

To analyze the behavior of the superfluid density for finite, non-zero values of $\alpha$, we developed a QMC algorithm targeting the power-law decay XY model. The SSE variant of QMC is a natural choice for computational studies targeting the superfluid density in $U(1)$ symmetric spin models \cite{sandvik2019Review2, melko2013SSEReview}. The superfluid density can be related to the winding number which can be extracted relatively easily in SSE by counting the number of off-diagonal operators in the resulting $(1+1)d$ configurations at each step of the simulation \cite{sandvik1997finite, melko2004aspect}. To validate our QMC approach, we used ED calculations with $N = 11$. Our ED algorithm, in turn, was validated by comparing the $\alpha \rightarrow \infty$ limit to the analytical FFS solution \cite{laflorencie2004scaling, laflorencie2001finite}. We also compared our QMC results to MFT and LMG results and found good agreement. Finally, to analytically verify the qualitative enhancement observed numerically using QMC, we supplemented our analysis with LSW theory \cite{coletta2012semiclassical, TommasoLSWLRXY, bernardet2002analytical} which is applicable to finite $\alpha > 0$. 

Using our QMC calculations, we showed that the superfluid density diverges for the long and medium range models, in line with expectations from bosonization theory results \cite{dupuis_2024, edmond_SCHA}. This divergence persists in the presence of the Kac normalization as well, which is typically used to control the super-extensivity of the ground state energy. We also analyzed the divergence free contribution to the superfluid density via a normalized superfluid density estimator, which is well-defined in the presence, as well as in the absence of the Kac normalization. This normalized superfluid density estimator clearly distinguishes the short, medium and long-range regimes. Our QMC results suggest a critical $\alpha_c \approx 2.7$ in line with perturbative field theory predictions \cite{maghrebi2017continuous}. While LSW yields the expected qualitative behavior, we found that it overestimates the normalized superfluid density for medium range interactions ($\alpha \in (1,3)$) and underestimates the expected results for the nearest-neighbor case. As $\alpha \rightarrow 0$, LSW and QMC results converge to the mean field values in the thermodynamic limit, and the spin wave correction reproduces the $\mathcal{O}(1/N)$ finite size scaling of the LMG results to the mean field predictions. 

The enhancement of superfluidity discussed in this work, while interesting from a theoretical perspective, also offers the potential for experimental realizations in trapped-ion quantum devices. Such architectures are uniquely suited to simulating the many-body physics of the long range XXZ and XY spin-$1/2$ models \cite{monroe2021programmable, teoh2020machine}. 
XXZ models may be particularly rich in phenomena, including regimes where long range order is possible in $1d$ \cite{1d_TLRO_Nahum, 1d_TLRO_Calderon}.
The coupling strengths $J_{ij}$ in trapped ion implementations are  determined by device control parameters and can be tuned to yield interaction strength distributions that are well-approximated by the $1/r^\alpha$ couplings studied here with $\alpha \in (0,3)$ \cite{richerme2014non}. Computational studies simulating the superfluid behavior of the native XY and XXZ trapped ion Hamiltonians will be explored in future work. 

\begin{acknowledgments}
We would like to thank Yi Hong Teoh, Ying-Jer Kao, Cenke Xu and Pierre-Nicholas Roy for insightful discussions. 
A.D. acknowledges support from the National Science Foundation QuSeC-TAQS program under award No. OSI-2326801.
We also acknowledge financial support from the Natural
Sciences and Engineering Research Council of Canada
(NSERC) and the Perimeter Institute. Research at
Perimeter Institute is supported in part by the Government of Canada through the Department of Innovation,
Science and Economic Development Canada and by the
Province of Ontario through the Ministry of Economic
Development, Job Creation and Trade.
\end{acknowledgments}

\section*{Data Availability}
All of the code developed and used for this work is publicly available (see Ref. \onlinecite{oqd_heisenberg_ion}). The data files that support the findings of this study are too large to host on GitHub, but can be made available upon reasonable request. 

\appendix

\section{Lipkin-Meshkov-Glick Model}

Here, we briefly review the Lipkin-Meshkov-Glick (LMG) model to derive the ground state energy and the superfluid density in the infinite range limit $(\alpha \rightarrow 0)$ of the XY model.

\subsection{Ground State Energy \label{sec:appendix_A_lmg_energy}}

The LMG Hamiltonian of interest is: 
\begin{gather}
    H = -\frac{J}{2}\sum_{i \neq j} (S_i^x S_j^x + S_i^y S_j^y). \label{appendix_alpha_0_untwisted_H}
\end{gather}
To diagonalize this, we first introduce the collective spin operators \cite{LMG2008}: 
\begin{equation}
    S_\gamma = \sum_{i=1}^N S_i^\gamma, \ \ \ \gamma \in \{x,y,z\}.
\end{equation}
These operators satisfy the usual $SU(2)$ algebra as can be seen by examining the commutator: 
\begin{equation}
    [S_u, S_v] = \sum_{i=1}^{N} [S^u_i, S^v_i] = i \varepsilon_{uvw} \sum_{i=1}^{N} S_i^w = i \varepsilon_{uvw} S_w, 
\end{equation}
where $u,v,w$ are in the ordered set $\{x,y,z\}$ and $\varepsilon_{uvw}$ is the Levi-Civita symbol \cite{sakurai2020modern}. The Hamiltonian can be expressed in terms of these operators as follows: 
\begin{equation}
    H = -\frac{J}{2} \left(S_x^2 + S_y^2 - \sum_{i=1}^{N} ((S_i^x)^2 + (S_i^y)^2) \right).
\end{equation}
The collective spin operators square to yield a sum over all pairs including self-interaction terms, which are subsequently removed by the sum over the squares of single site spin operators. Note that $(S_i^\gamma)^2 = 1/4$ for all $\gamma \in \{x,y,z\}$. Therefore, the Hamiltonian can be entirely expressed in terms of the collective spin operators as follows: 
\begin{equation}
    H = -\frac{J}{2} (S_x^2 + S_y^2) + \frac{JN}{4} = -\frac{J}{2} (S^2 - S_z^2) + \frac{JN}{4}. \label{appendix_collective_spin_H}
\end{equation}

Note that $S^2$ above is the $SU(2)$ Casimir:  $S_x^2 + S_y^2 + S_z^2$. This Hamiltonian can be diagonalized using the spin states: $|E(S,M) \rangle = |S,M\rangle$ with $S$ either an integer or half integer and $M \in \{-S, -S+1, ... , S-1, S\}$ for a given $S$ \cite{TommasoLSWLRXY, Latorre2005LMG}. The energy for a given set of quantum numbers is given by: 
\begin{equation}
    E(S,M) = -\frac{J}{2} (S (S+1) - M^2) + \frac{JN}{4}.
\end{equation}

To determine the ground state, we can minimize the above with respect to $S$ and $M$ which corresponds to maximizing $S$ and minimizing $M$. Note that $M$ is the eigenvalue of the $S_z$ operator and $S_z$ is the cumulative magnetization over $N$ sites. Therefore, $M$ is bounded above by $N/2$ and the largest possible spin quantum number in the many-body Hilbert space is $S_{max} = N/2$. It is easy to show that $(S=N/2, M=1/2)$ minimizes the energy for any odd $N > 0$. The ground state energy is then given by: 
\begin{equation}
    E_0 = -\frac{J}{8} (N^2 - 1). \label{appendix_energy_gs_lmg}
\end{equation}

\subsection{Superfluid density \label{sec:appendix_A_lmg_stiffness}} 

To determine $\rho_s$, we need to analyze the twisted Hamiltonian: 
\begin{gather}
    H = -\frac{J}{2} \sum_{i \neq j} \biggl ( \cos(r_{ij} \theta) (S_{i}^x S_j^x + S_i^y S_j^y) \nonumber \\ + \sin(r_{ij} \theta) (S_{i}^x S_{j}^x - S_{i}^y S_{j}^y) \biggr). \label{appendix_twisted_lgm_H}
\end{gather}
The second term with coefficient $\sin(r_{ij} \theta)$ is the current density $j_b$, which has a trivial expectation value in the ground state: $\langle j_b \rangle = 0$ \cite{sandvik1997finite}. We can approximate $\rho_s$ using finite differences as follows: 
\begin{equation}
    \rho_s = \lim_{\theta \rightarrow 0} \frac{2}{N} \frac{\langle H(\theta) \rangle  - \langle H(0) \rangle }{\theta^2}. \label{appendix_rho_s_finite_differences_def}
\end{equation}
Here, the expectation values are evaluated in the ground state. Therefore, it suffices to only consider the first term in Eq. \eqref{appendix_twisted_lgm_H}. Note that the collective spin operator can not be directly defined as for the untwisted case here because of the bond-dependent coupling $\cos(r_{ij} \theta)$. However, since we are interested in the limit of $\theta \rightarrow 0$, we can restrict to $\theta^2 \ll (N-1)^2 \Rightarrow \cos(r_{ij} \theta) \approx 1 - r_{ij}^2 \theta^2/2$ and analyze the effect of the twist perturbatively. To this end, note that the first order ground state energy shift is given by: 
\begin{equation}
    \Delta E_0 = \frac{J \theta^2}{4} \sum_{i \neq j} r_{ij}^2 \langle S_{i}^x S_j^x + S_i^y S_j^y \rangle_{\theta = 0}. \label{appendix_Delta_E_lmg}
\end{equation}

The expectation value needs to be evaluated in the unperturbed ground state determined by the Hamiltonian in Eq. \eqref{appendix_alpha_0_untwisted_H}. Since all bonds correspond to the same interaction strength in Eq. \eqref{appendix_alpha_0_untwisted_H}, we can assume that the ground state expectation value of each local bond Hamiltonian $H_b = \langle S_{i(b)}^x S_{j(b)}^x + S_{i(b)}^y S_{j(b)}^y\rangle_{\theta=0}$ does not depend on the bond $b$. Factoring the expectation value yields for the energy shift: 
\begin{equation}
    \Delta E_0 = \langle H_b \rangle_{\theta=0} \frac{J \theta^2}{4} \sum_{i \neq j} r_{ij}^2.
\end{equation}

The expectation value $\langle H_b \rangle_{\theta=0}$ can be computed by equally distributing the collective spin energy contribution across all bonds as follows: 
\begin{gather}
    \langle H_b \rangle_{\theta=0} = \frac{1}{N(N-1)} \left \langle \sum_{i \neq j}  S_{i}^x S_j^x + S_i^y S_j^y \right \rangle \nonumber \\ =  - \frac{2}{J N(N-1)} \langle H(\theta=0) \rangle = \frac{1}{4} \frac{(N^2 - 1)}{N(N-1)}.
\end{gather}
Here, in the final equality we have used Eq. \eqref{appendix_energy_gs_lmg} to evaluate the expectation value of the untwisted Hamiltonian. The sum over squares in Eq. \eqref{appendix_Delta_E_lmg} can be evaluated analytically as well \cite{graham_concrete}: 
\begin{equation}
    \sum_{i \neq j} r_{ij}^2 = 2N \sum_{r=1}^{(N-1)/2} r^2 = \frac{N^2(N^2-1)}{12}. \label{appendix_sum_square_distances}
\end{equation}
This finally gives for the superfluid density: 
\begin{equation}
    \rho_s = \frac{J(N+1)}{8} \left (\frac{N^2 - 1}{12} \right). \label{appendix_lmg_stiffness}
\end{equation}

\section{\label{sec:appendix_B_mft} Mean Field Solution}
To determine the mean field behavior, consider again the Hamiltonian of the uniformly interacting model: 
\begin{gather}
    H_0 = -\frac{1}{2}\sum_{i \neq j} J_{ij} (S_i^x S_j^x + S_i^y S_j^y) \nonumber \\ = -\frac{1}{2} \sum_{i \neq j} J_{ij} \ \left(\vec{S_i} \cdot \vec{S_{j}} \right).
\end{gather}
Here, we have simply collected the in-plane spin components into vectors: $\vec{S_i} = \left(S_i^x, S_i^y\right)$. Now, adding an external field in the $X$ direction without loss of generality (any direction in the $XY$ plane would suffice because of the $U(1)$ symmetry) \cite{coletta2012semiclassical}, we get:
\begin{equation}
    H = H_0 + H_1 = -\frac{1}{2} \sum_{i \neq j} J_{ij} \ \left(\vec{S_i} \cdot \vec{S_{j}} \right) - h\sum_{i} S_{i}^x.
\end{equation}

We will analyze the behavior of this system in the limit $h \rightarrow 0$ to recover our original system of interest. Note that the direction of the mean field is defined by the direction of the external field $\langle \vec{S} \rangle = \langle \left(S_x \rangle, 0\right) \equiv (X, 0)$.  Ignoring fluctuations about the mean field: $(\vec{S_{i}} - \langle \vec{S} \rangle) (\vec{S_{j}} - \langle \vec{S} \rangle) \approx 0$ \cite{goldenfeld2018lectures}, our Hamiltonian reduces to: 
\begin{equation}
    H = -\tilde{J}X \sum_{j} S_{j}^x + \frac{1}{2} \tilde{J}N X^2 - h\sum_{j} S_j^x. \label{appendix_mft_hamiltonian}
\end{equation}
We have defined $J_i = \sum_{j \neq i} J_{ij}$ and used the isotropy of space: $J_{i} = J_{j} \equiv \tilde{J}$, in the above equation. The Hamiltonian in Eq. \eqref{appendix_mft_hamiltonian} yields the usual mean field equation for the order parameter which can be obtained by minimizing the free energy per particle with respect to $X$: 
\begin{equation}
    X = \frac{1}{2} \tanh\left(\frac{\beta}{2} (\tilde{J}X + h)\right).
\end{equation}

As $\beta \rightarrow \infty$, this yields $X = \frac{1}{2}$. As before, we now analyze the twisted Hamiltonian to determine $\rho_s$ ignoring the current term due to its trivial expectation value in the ground state: 
\begin{gather}
    H_0 = -\frac{1}{2} \sum_{i \neq j} J_{ij} \cos(r_{ij} \theta) (S_{i}^x S_j^x + S_i^y S_j^y).
\end{gather}

Defining $J_{ij}(\phi) = J_{ij} \cos(r_{ij} \phi)$, we get the same form for the mean field ground state energy as the untwisted case. To get the mean field solution for $\rho_s$ using finite differences, we can evaluate the expectation values of $H(\theta)$ and $H(0)$ in the mean field ground state corresponding to the limit $h \rightarrow 0^{+}$. This is given by $|+\rangle^{N}$ \cite{TommasoLSWLRXY, bernardet2002analytical} which yields the following for the superfluid density: 
\begin{gather}
    \rho_s = -\frac{\Delta \tilde{J} X^2}{N\theta^2} = -\frac{X^2}{N\theta^2} \sum_{j \neq i} J_{ij} (\cos(r_{ij} \theta) - 1) \\ \Rightarrow \rho_s \approx \frac{JX^2}{2N} \sum_{j \neq i} r_{ij}^2 = \frac{J}{4} \sum_{r=1}^{(N-1)/2} r^2.
\end{gather}
In the last equality above, we have used the mean field solution in the zero-temperature limit: $X = \pm \frac{1}{2}$ after expanding $\cos(r_{ij} \theta) \approx 1 - r_{ij}^2 \theta^2/2$. Evaluating the sum using Eq. \eqref{appendix_sum_square_distances}, we get for the superfluid density:
\begin{equation}
    \rho_s = \frac{JN}{8} \left(\frac{N^2-1}{12} \right). \label{appendix_mft_rho_s}
\end{equation}
Similarly, the mean field ground state energy can be determined using Eq. \eqref{appendix_mft_hamiltonian}: 
\begin{equation}
    E_0 = -\frac{JN (N-1)}{8}. \label{appendix_mft_energy}
\end{equation}

\section{Linear Spin Wave Theory}

\subsection{Construction \& Diagonalization \label{sec:appendix_C_lsw_diagonalization}}

To construct the LSW theory, consider the power-law decay XY model Hamiltonian in the presence of a twist: 
\begin{equation}
    H(\theta) = -\frac{J}{2} \sum_{i=0}^{N-1} \sum_{r \neq 0} \frac{\cos(r \theta)}{|r|^\alpha} (S_i^x S_{i+r}^x + S_i^y S_{i+r}^y ). \label{appendix_lsw_Hamiltonian_1}
\end{equation}
We have symmetrized the sum from Eq. \eqref{sum_pairs_H}, and defined $j = i+r$ with $r \in [-(N-1)/2, (N-1)/2]$, $r \neq 0$ for the periodic chain with $N$ odd, in the above equation. We have also neglected the current term associated with a non-zero twist in the Hamiltonian since it does not contribute to the ground state energy, as before. As is typical for the nearest neighbor XY model, we proceed by mapping our spin operators to Holstein-Primakoff bosons as follows \cite{HolsteinPrimakoffBosons}: 
\begin{gather}
    S^x_i \approx \frac{1}{2} - a_i^\dagger a_i, \\
    S^y_i \approx \frac{1}{2}(a_i + a_i^\dagger).
\end{gather}

It is instructive to note that $a_i^\dagger$ and $a_i$ now represent boson raising and lowering operators respectively, and should not be confused with the hardcore bosons discussed in Sec. \ref{sec:stiffness} (denoted by $b_i^\dagger$ and $b_i$). The resulting Hamiltonian is: 
\begin{gather}
    H(\theta) = -\frac{J}{2} \sum_{i} \sum_{r \neq 0} \frac{\cos(r \theta)}{|r|^\alpha} \biggl[ \frac{1}{4} - \frac{1}{2} ( a_i^\dagger a_i + a_{i+r}^\dagger a_{i+r}) \nonumber \\ + \frac{1}{4} (a_{i} a_{i+r} + a_{i}^\dagger a_{i+r}^\dagger + a_{i} a_{i+r}^\dagger + a_{i}^\dagger a_{i+r}) \biggr], \label{appendix_lsw_Hamiltonian_2}
\end{gather}
where we have neglected the quartic term resulting from the product $S_y^i S_y^{i+r}$ in Eq. \eqref{appendix_lsw_Hamiltonian_1} to obtain a linear approximation for spin wave theory \cite{coletta2012semiclassical}. Higher order terms correspond to interactions between the spin waves which are ignored by construction in LSW theory. 

To diagonalize the resulting approximate Hamiltonian, we first Fourier transform Eq. \eqref{appendix_lsw_Hamiltonian_2}: 
\begin{gather}
    H(\theta) = -\frac{JN \gamma_0 (\theta)}{4} + \sum_k A_k(\theta) (a_k^\dagger a_k + a_{-k}^\dagger a_{-k}) \nonumber \\ - \sum_{k} B_k(\theta) (a_k^\dagger a_k + a_{-k}^\dagger a_{-k}), \label{appendix_FT_lsw_H}
\end{gather}
where $a_k = N^{-1/2}\sum_j e^{-i 2 \pi jk/N} a_j$ and $a_k^\dagger = N^{-1/2}\sum_j e^{i 2 \pi jk/N} a_j^\dagger$ are the Fourier transforms of $a_j$ and $a_j^\dagger$ respectively. The coefficients $A_k$ and $B_k$ are given by: 
\begin{gather}
    A_k(\theta) = \frac{J}{2} \left(\gamma_0(\theta) - \frac{\gamma_k(\theta)}{2} \right), \ \ B_k = \frac{J \gamma_k(\theta)}{4}, \\ \gamma_k(\theta) = \sum_{r=1}^{(N-1)/2} \frac{\cos(2 \pi r k/N) \cos(r \theta)}{r^\alpha}.  \label{appendix_gamma_k_full_def}
\end{gather}

Note that the form in Eq. \eqref{appendix_FT_lsw_H} is amenable to Bogoliubov diagonalization \cite{bogoliubov1947theory}. To this end we define: 
\begin{gather}
    a_k = u_k \xi_k + v_k \xi_{-k}^\dagger, \\
    a_k^\dagger = u_k \xi_k^\dagger + v_k \xi_{-k},
\end{gather}
where $\xi_k$ and $\xi_k^\dagger$ are the quasi-particle raising and lowering operators. Imposing the bosonic commutation relation constrains the coefficients $u_k$ and $v_k$ to be hyperbolic: $u_k^2 - v_k^2 = 1$. Expressing the Hamiltonian in Eq. \eqref{appendix_FT_lsw_H} in terms of these Bogoliubov quasi-particles, we get the following diagonalized form \cite{TommasoLSWLRXY}: 
\begin{gather}
    H = -\frac{JN \gamma_0}{4} + \sum_k \left(\sqrt{A_k^2 - B_k^2} \right) (\xi_k^\dagger \xi_k + \xi_{-k}^\dagger \xi_{-k}) \nonumber \\ + \sum_k \left(\sqrt{A_k^2 - B_k^2} - A_k\right),
\end{gather}
where we have suppressed the $\theta$ dependence for brevity. The energy of each quasi-particle mode is given by the following: 
\begin{equation}
    \epsilon_k = 2 \left(\sqrt{A_k^2 - B_k^2} \right) = J \gamma_0 \left(1 - \frac{\gamma_k}{\gamma_0}\right)^{1/2}.
\end{equation}

The ground state corresponds to the vacuum state (no quasi-particles) and the ground state energy is given by: 
\begin{equation}
    E_0 = -\frac{JN \gamma_0}{4} + \sum_k \left(\sqrt{A_k^2 - B_k^2} - A_k\right). \label{appendix_lsw_energy}
\end{equation}
The first term in the above expression is typically referred to as the classical solution and the second term is the LSW correction \cite{coletta2012semiclassical, bernardet2002analytical}. 

\subsection{Mean Field Limit \label{sec:appendix_C_lsw_alpha_0}}

Here, we derive the ground state energy and the superfluid density for $\alpha=0$ in LSW theory analytically. The ground state energy can be obtained by setting $\theta=0$. For these limits ($\alpha = \theta = 0$), $\gamma_0 = (N-1)/2$. Therefore, the first term of the ground state energy in Eq. \eqref{appendix_lsw_energy} recovers exactly the mean field result from Eq. \eqref{appendix_mft_energy}. Moreover, the series corresponding to $\gamma_k$ for $k \neq 0$ can be summed analytically since it corresponds to the Dirichlet kernel \cite{rudin1987real}. For $\theta = 0$, we get: 
\begin{gather}
    \gamma_k = \sum_{r=1}^{(N-1)/2} \cos(2 \pi r k/N) = -\frac{1}{2}.
\end{gather}

The LSW correction term in Eq. \eqref{lsw_energy} is given by the following: 
\begin{gather}
    \Delta E_0 = -\frac{J \gamma_0}{2} \biggl(N  - \frac{1}{2} \biggr) \nonumber \\ -\frac{J \gamma_0}{2} \Biggl(\sum_{k=1}^{N-1} \biggl[ \left(1 + \frac{1}{2\gamma_0}\right)^{1/2} - \frac{1}{4 \gamma_0} \biggr] \Biggr).
\end{gather}

We can binomially expand the expression above since the thermodynamic limit corresponds to $\gamma_0 \rightarrow \infty$, to get the LSW estimate for the ground state energy: 
\begin{gather}
    E_0^{LSW} = -\frac{J\gamma_0(N+1)}{4}, \label{E_0_eqn_lsw} \\ \Rightarrow E_0^{LSW} = -\frac{J(N^2-1)}{8}.
\end{gather}

To analyze the superfluid density associated with the $\alpha=0$ limit using LSW theory, we need to evaluate the ground state energy in the presence of a small twist $\theta$. Restricting $\theta \ll 1/N$, first note that $\gamma_k/\gamma_0 \sim O(1/N) \rightarrow 0$ as $N \rightarrow \infty$. This can be explicitly shown by using the Dirichlet kernel to analytically sum the series' $\gamma_0$ and $\gamma_k$ for $k \neq 0$ and then evaluating the quotient series $\gamma_k/\gamma_0$. Therefore, for large $N$, the energy shift can be approximated binomially, as before, for $k \neq 0$ and evaluated directly for $k=0$. This again yields Eq. \eqref{E_0_eqn_lsw} for the LSW estimate of $E(\theta)$, with $\gamma_0$ in this case corresponding to $\gamma_0(\theta)$. Evaluating the superfluid density using finite differences gives: 
\begin{equation}
    \rho_s^{LSW} = \frac{J(N+1)}{8} \left (\frac{N^2 - 1}{12} \right). 
\end{equation}

Finally, it is also interesting to consider the energy of the quasi-particle modes for $\alpha = 0$: 
\begin{equation}
    \epsilon_k =  \frac{J(N-1)}{2} \left(\frac{N}{N-1}\right)^{1/2}. \label{epsilon_k_alpha_0}
\end{equation}
The excitation energies are independent of $k$ and scale linearly with the number of sites $N$. As discussed in the main text, this causes QMC calculations at high temperatures to converge to ground state results as $\alpha \rightarrow 0$. 


%

\end{document}